\documentclass[a4paper,fleqn,usenatbib]{mnras}

\usepackage[varg]{txfonts}

\usepackage{bm}
\usepackage{graphicx}
\usepackage{amssymb}	
\usepackage{subfigure}
\usepackage{color}

\def\rd{\mathrm{d}}

\def\au{\,\mathrm{a.u.}}
\def\cm1{\,\mathrm{cm}^{-1}}

\def\Ry{\,\mathrm{Ry}}
\def\kelvin{\,\mathrm{K}}

\title[EIE Ni$^{3+}$ and PI Ni$^{2+}$]
      {Spectroscopic diagnostics of low-ionized iron-peak elements.
       Electron-impact excitation of Ni$^{3+}$ and 
       photoionization of Ni$^{2+}$.}
\author[L. Fern\'{a}ndez-Menchero et al.]{
        L. Fern\'{a}ndez-Menchero, 
        R. T. Smyth, 
        C. A. Ramsbottom and
        C. P. Ballance  
        \\
        Centre of Theoretical Atomic, Molecular and Optical Physics.
        Queen's University Belfast. \\
        University Road, Belfast, BT7 1NN, Northern Ireland, United Kingdom
        }
\date{Accepted XXX. Received YYY; in original form ZZZ}
\pubyear{2017}

\begin{document}
\label{firstpage}
\pagerange{\pageref{firstpage}--\pageref{lastpage}}
\maketitle

\begin{abstract}

The spectra from Fe-peak elements may be used to determine the temperature 
and density of various astrophysical objects. 
Determination of these quantities is underpinned by the accuracy 
and the comprehensiveness of the underlying atomic structure and collisional 
calculations. 
In the following paper, we shall focus specifically on \ion{Ni}{iv} lines 
associated with transitions amongst several low-lying levels.
We shall employ modified versions of the parallel Dirac R-matrix codes, 
considering both electron-impact excitation of Ni$^{3+}$ and the photoionisation 
of both the ground and excited states of Ni$^{2+}$.
We produce high-quality data sets for both processes, 
and using these data, we calculate line ratios relevant for plasma diagnostics 
of temperature and density.

\end{abstract}

\begin{keywords}
Atomic data -- opacity -- photoionization -- electron-impact excitation
\end{keywords}


\section{Introduction}
\label{sec:introduction}

The spectra of lowly ionized iron-peak elements such as
($\mathrm{Fe}^{q+}$, $\mathrm{Co}^{q+}$, $\mathrm{Ni}^{q+}$, $q=0-3$)
are vitally important in astronomical observation.
In particular, $\mathrm{Ni}^{3+}$ being iso-electronic with Fe$^{+}$, 
should produce many of the same diagnostic lines, 
which have extensively been studied previously by 
\citet{pradhan1993,zhang1995,ramsbottom2005,ramsbottom2007,ramsbottom2009}.
Comparisons of these lines using the results of the present paper shall be made.
The Fe-peak elements provide some of the most abundant species created inside 
the stars, and they emit at ultraviolet wavelengths, making them dominant 
contributors to the opacity of the interstellar media under certain conditions.

From an atomic physics perspective, the half-open d-shell nature of 
many of these systems inevitably leads to target descriptions involving 
between twenty and thirty configurations if spectroscopic accuracy is 
to be approached.
The $N+1$-electron collisional calculation, whether it be excitation 
or photoionisation expands to target descriptions involving  
between five- and seven-thousand levels with the associated cost of 
calculating over $10^{9}$ Racah angular coefficients. 
Only with the development of the current suite of codes, including multiple 
layers of hyper-threaded parallelism, 
the Hamiltonian formation being the most critical part, 
have we been able to make progress on these type of systems. 
It has enabled us to provide comprehensive data sets that include {\it every} 
excitation and de-excitation for electron-impact excitation (EIE) 
or photoionisation (PI), 
not just from the ground state, but as well from {\it every} excited state.

Several groups throughout the world \citep{opacity1995}%
\footnote{\url{http://cdsweb.u-strasbg.fr/topbase/TheOP.html}}, 
OPAL \citep{iglesias1996}\footnote{\url{https://opalopacity.llnl.gov/}}, 
recognise that calculated opacities are essential for the correct interpretation 
of the spectra taken from a variety of objects, such as interstellar clouds, 
nebulae, remains of supernovae and stellar atmospheres. 
Even today, there are still remaining outstanding issues with experimental 
measurements that do not agree with any of the predicted models listed 
above \citep{bailey2015}. 
We note that fundamental atomic data is only one aspect of 
the complex plasma-modelling codes, 
but the ability to put a realistic uncertainty on first principle calculations 
can eliminate it as the cause of disagreement with experimental observations. 
This has spurred the calculation of uncertainties with every collisional process, 
that are both a function of temperature and density. 
Our long-term goal is to ensure that the astrophysical modelling packages 
such as {\sc cloudy} \citep{ferland2017}\footnote{\url{https://www.nublado.org}}, 
Xstar\footnote{\url{https://heasarc.nasa.gov/lheasoft/xstar}}
can integrate not only new atomic collisional data but also the associated 
uncertainty file.

One long-term project that recognised the need for comprehensive photoionisation 
of every ion stage for a large part of the periodic table is 
the Opacity Project~\citep{opacity1995}. 
The distribution of work on various atomic species was calculated by theoretical 
physics groups across the world and provided a very fruitful collaboration.
However, as the near-neutral species are highly complex and 
the former computational resources were insufficient to calculate them, 
the Opacity Project saw a greater focus on the more highly ionised systems. 
With the exception of iron, and several large scale LS resolved models,
comprehensive level resolved calculations involving several hundred 
states remain to be calculated for the near-neutral Fe-peak elements.     

Currently within the literature, some of the following works represent 
historical attempts at calculating lowly-ionized iron peak elements, 
namely \citet{pradhan1993,zhang1995} calculated the EIE 
of $\mathrm{Mn}$-like $\mathrm{Fe^{+}}$ in an $LS$ coupling formalism.
Their close-coupling expansion employed only three configurations,
resulting in 38 $LS$ terms, enabling transitions only among the ground and first 
excited configurations. 
We appreciate the limitations of this small model, and also that it has taken 
another twenty years to include twenty more configurations in the 
configuration-interaction (CI) description of our present model.
\citet{ramsbottom2005,ramsbottom2007,ramsbottom2009} 
also within an LS-coupling framework made a succession of calculations
with a progressively better target description. 
They ultimately included orbitals up to $n=4$ resulting in a  total of 113 $LS$
terms in their close coupling (CC) expansion.
Other set of works for other isoelectronic sequences of low-ionized iron peak 
elements includes the one of \citet{zhang1997} for EIE of $\mathrm{Fe}^{3+}$, 
and \citet{bautista2004} for EIE of~$\mathrm{Ni^{+}}$.

Although not as dominant as iron, nickel lines are also used for diagnostic
and modeling of astrophysical plasmas.
\citet{mazzali2001} performed several models for type Ia supernovae concluding 
that nickel abundance can affect its brightness and decline rate.
Years later, nickel lines were observed in the remnant of the supernova 1987A 
by \citet{mccray2016}.
Furthermore, \citet{werner2018} subsequently used the absorption features in 
white dwarf atmospheres produced by iron-peak elements to model the metal abundances.
Opacity data are necessary for any kind of simulation work so the demand for 
comprehensive data sets is almost insatiable.
One such example of these simulations is the work of \citet{sanchez2007}.
They used models dependent upon opacities and the known optical depths of 
the interstellar clouds to determine its fractal dimension.
Another example is the work of \citet{moravveji2016}, whose opacity simulations, 
comparing measured to measured spectra, concluded that nickel ions produce an 
enhancement of the opacity.

Opacity is also an important aspect of the {\sc cloudy} software 
package \citep{ferland2017}. 
{\sc cloudy} is extensively used for the simulation of the spectra collected from 
interstellar clouds.

For completeness, the present work will investigate the $\mathrm{Mn}$-like 
ion $\mathrm{Ni^{3+}}$ for two important processes: 
the EIE and the PI of its parent ion $\mathrm{Ni^{2+}}$.
We employ a heavily modified parallel version of the fully-relativistic 
Dirac Atomic R-matrix code ({\sc darc}) 
\citep{norrington1987,ballance2004}\footnote{http://connorb.freeshell.org}. 
We include 23 configurations in the CI expansion.
This expansion leads to a total of $6\,841$ relativistic levels.
From that total, we reduce the CC expansion to include the first $262$ levels,
this reduction of the basis set may lead to pseudoresonances,
and we have to take in account this fact when analysing the final collision strengths.
$\mathrm{Ni^{3+}}$ is a low-ionized intermediate-mass ion, 
therefore there is the expectation that relativistic effects will not be large,
especially for valence-shell electrons.
One might argue on theoretical grounds that a semi-relativistic formalism, 
or even a non-relativistic one, 
would lead to acceptable results with considerably less computational effort.
However, the current multi-level parallelism of the parallel {\sc darc} suite
of codes, whilst being more computationally intensive is currently 
considerably more efficient than Breit-Pauli or ICFT semi-relativistic 
versions. 

Over the last few years considerable effort has been made by the group 
at Queen's University Belfast in refactoring codes,
specifically in terms of memory management. 
The last versions of the {\sc darc} code are viable, factoring in hardware 
limitations, to handle thousands of target states in the CC expansion.

The remainder of the paper is organized as follows:
in Section~\ref{sec:structure} we give our description of the atomic structure;
in Section~\ref{sec:scattering} we describe the close-coupling method used to 
obtain the EIE collision strengths and subsequent effective collision strengths 
as well as the PI cross sections;
in Section~\ref{sec:results} we show and discuss the results;
in Section~\ref{sec:modeling} we perform a simple collision-radiative model 
to test the diagnostics predicted with present collision rates in relation 
to \ion{Fe}{ii} work;
and in Section~\ref{sec:conclusions} we discuss the conclusions of the work.
Atomic units are used unless otherwise specified.

\section{Structure}
\label{sec:structure}

We use the General-purpose Relativistic Atomic Structure Package ({\sc grasp}) 
\citep{dyal1989,parpia1996} to determine the best possible atomic structure 
within a Dirac-Coulomb framework. 
The resulting radial orbitals from this Multi-Configuration Dirac-Fock (MCDF) 
method are defined on an exponential radial grid and they are employed 
subsequently in the electron-impact excitation calculation. 

In our CI expansion we permute the 25 electrons of the $\mathrm{Mn}$-like 
$\mathrm{Ni}$ target within the configurations given below.
Thirteen non-relativistic orbitals, namely the 
1s, 2s, 2p, 3s, 3p, 3d, 4s, 4p, 4d, 5s, 5p, 6s, 6p
are transformed into their relativistic counterparts within {\sc grasp}.
To optimise the CI expansion and to accelerate the MCDF process we follow 
several steps, validating our results against the recommended values of the 
NIST atomic spectra data table \citep{nist2018,sugar1985} where available.
In our first step, we included the ground state 
configuration $\mathrm{Ne\,3s^2\,3p^6\,3d^6\,4s}$,
and all possible one-electron excitations $\mathrm{3s^2\,3p^6\,3d^6}\,nl$.
This simple expansion led to a first approximation of the one-electron 
wave functions.
Additional configurations only slightly refine the core orbitals up to the 
3s, but do help the convergence of the valence orbitals. 
With each iteration we check the updated excitation energies of the first $50$ levels
with the recommended data of NIST, with the goal of a compact but accurate basis. 
The results of our final 23 configuration model are listed in Table~\ref{tab:confs},
though here we only provide a representative sample of the possible $6\,841$ 
relativistic levels, the supplementary online material shall be more comprehensive. 
Comparing our calculated excitation energies with respect to the ground level with 
the recommended values of NIST we find our largest deviation in the order $12\%$,
and an average deviation of $3.3\%$.
This deviation is quite acceptable 
considering the complexity of the system and comparisons with the previous works 
for $\mathrm{Mn}$-like $\mathrm{Fe}$ of \citet{ramsbottom2005}, 
whose largest deviation was order $15\%$ in $LS$ coupling,
and the one of \citet{pradhan1993}, order~$25\%$.  

For a further comparison and to quantify the uncertainty in the atomic structure 
we performed a second independent calculation using a different atomic structure code.
The {\sc autostructure} program \citep{badnell2011b} code serves this purpose.
{\sc Autostructure} provides non-relativistic radial wave functions from a 
Thomas-Fermi-Amaldi potential for the 1s to 6p orbitals.
The subsequent Breit-Pauli Hamiltonian includes the relativistic terms as a first
order perturbations: mass-velocity, spin-orbit and Darwin.
We neglect the second order perturbation terms spin-spin, orbit-orbit and 
spin-other-orbit.
To determine the $\lambda_{nl}$, or scaling parameters within the TFA model 
potential, we variationally determine them from minimization of the absolute 
Hamiltonian energy.  
For a balanced comparison with the {\sc grasp} and {\sc darc} calculations
and in order to minimise the differences in atomic structure, 
we keep both problems as similar as possible.
In that regard, we include in the CI expansion of the {\sc autostructure} model 
exactly the same configuration set as that in {\sc grasp}.
After performing the minimization process we obtained the values of $\lambda_{nl}$ 
shown in Table~\ref{tab:lambda}.
In the supplementary online material we show the energies obtained with 
{\sc autostructure} for a complete comparison with the ones obtained with {\sc grasp} 
and in Table~\ref{tab:energies} we show here the lowest-energy $50$ levels.
The level energies obtained with {\sc autostructure} deviate slightly further from 
the recommended values of NIST than the ones obtained with {\sc grasp}.
The maximum deviation is of the order of $15\%$, again larger than the {\sc grasp} one.

\begin{table}
\caption{Configuration list included in the atomic structure calculations}
\label{tab:confs}
\centering
\begin{tabular}{l@{\qquad}l@{\quad}|@{\quad}l@{\quad}}
  \hline
  \multicolumn{2}{l@{\quad}|@{\quad}}{Even parity} & Odd parity \\
  \hline
  \multicolumn{3}{l}{Core: $\mathrm{1s^2\,2s^2\,2p^6\,3s^2}$}   \\
  \hline
  $\mathrm{3p^6\,3d^6\,4s  }$ & $\mathrm{3p^5\,3d^7\,4p  }$ &  
  $\mathrm{3p^6\,3d^5\,4s\,4p}$ \\
  $\mathrm{3p^6\,3d^7      }$ & $\mathrm{3p^5\,3d^7\,5p  }$ &  
  $\mathrm{3p^6\,3d^6\,4p    }$ \\
  $\mathrm{3p^6\,3d^5\,4s^2}$ & $\mathrm{3p^4\,3d^7\,4s^2}$ &  
  $\mathrm{3p^6\,3d^6\,5p    }$ \\
  $\mathrm{3p^6\,3d^5\,4p^2}$ & $\mathrm{3p^5\,3d^7\,6p  }$ &  
  $\mathrm{3p^6\,3d^6\,6p    }$ \\
  $\mathrm{3p^6\,3d^6\,5s  }$ & $\mathrm{3p^6\,3d^5\,4d^2}$ &  
  $\mathrm{3p^5\,3d^7\,4s    }$ \\
  $\mathrm{3p^6\,3d^5\,5s^2}$ & $\mathrm{3p^6\,3d^6\,4d  }$ &  
  $\mathrm{3p^5\,3d^6\,4s^2  }$ \\
  $\mathrm{3p^6\,3d^5\,5p^2}$ & $\mathrm{3p^4\,3d^9      }$ &  
  $\mathrm{3p^5\,3d^7\,5s    }$ \\
  $\mathrm{3p^6\,3d^6\,6s  }$ &                             &  
  $\mathrm{3p^5\,3d^7\,6s    }$ \\ 
  \hline
\end{tabular}
\end{table}

\begin{table}
\caption{Scaling parameters optimised by {\sc autostructure}}
\label{tab:lambda}
\centering
\begin{tabular}{ll@{\quad}|@{\quad}ll}
  \hline
  1s & $1.42396$ & 4s & $1.04299$ \\
  2s & $1.30959$ & 4p & $1.04410$ \\
  2p & $1.12342$ & 4d & $1.55730$ \\
  3s & $1.10133$ & 5s & $1.07620$ \\
  3p & $1.06211$ & 5p & $1.03593$ \\
  3d & $1.04845$ & 6s & $1.02930$ \\
     &           & 6p & $1.01296$ \\
  \hline
\end{tabular}
\end{table}

\begin{table*}
\caption{Excitation energies of the first 50 $\mathrm{Ni^{3+}}$ target levels 
   included in the present calculations}
\label{tab:energies}
\begin{tabular}{rllrlrrrrr}
\hline
 $  i$ & Configuration                      & Term             & $J$  & parity & GRASP      & AS         & NIST       & Err GRASP (\%)  & Err AS (\%)  \\
\hline
 $  1$ & $\mathrm{3p^6\,3d^7}$              & $\mathrm{^4F}$   & $9/2 $ & even & $     0.0$ & $     0.0$ & $     0.0$ & $   - $ & $   - $ \\
 $  2$ & $\mathrm{3p^6\,3d^7}$              & $\mathrm{^4F}$   & $7/2 $ & even & $  1094.5$ & $  1210.7$ & $  1189.7$ & $ -8.0$ & $  1.8$ \\
 $  3$ & $\mathrm{3p^6\,3d^7}$              & $\mathrm{^4F}$   & $5/2 $ & even & $  1889.7$ & $  2083.7$ & $  2042.5$ & $ -7.5$ & $  2.0$ \\
 $  4$ & $\mathrm{3p^6\,3d^7}$              & $\mathrm{^4F}$   & $3/2 $ & even & $  2431.3$ & $  2675.5$ & $  2621.1$ & $ -7.2$ & $  2.1$ \\
 $  5$ & $\mathrm{3p^6\,3d^7}$              & $\mathrm{^4P}$   & $5/2 $ & even & $ 18113.0$ & $ 19394.3$ & $ 18118.6$ & $  0.0$ & $  7.0$ \\
 $  6$ & $\mathrm{3p^6\,3d^7}$              & $\mathrm{^4P}$   & $3/2 $ & even & $ 18459.6$ & $ 19720.9$ & $ 18366.8$ & $  0.5$ & $  7.4$ \\
 $  7$ & $\mathrm{3p^6\,3d^7}$              & $\mathrm{^4P}$   & $1/2 $ & even & $ 18956.7$ & $ 20317.5$ & $ 18958.4$ & $  0.0$ & $  7.2$ \\
 $  8$ & $\mathrm{3p^6\,3d^7}$              & $\mathrm{^2G}$   & $9/2 $ & even & $ 21941.1$ & $ 22190.3$ & $ 19829.6$ & $ 10.6$ & $ 11.9$ \\
 $  9$ & $\mathrm{3p^6\,3d^7}$              & $\mathrm{^2G}$   & $7/2 $ & even & $ 22987.4$ & $ 23331.6$ & $ 20947.6$ & $  9.7$ & $ 11.4$ \\
 $ 10$ & $\mathrm{3p^6\,3d^7}$              & $\mathrm{^2P}$   & $3/2 $ & even & $ 25818.1$ & $ 26220.2$ & $ 23648.9$ & $  9.2$ & $ 10.9$ \\
 $ 11$ & $\mathrm{3p^6\,3d^7}$              & $\mathrm{^2P}$   & $1/2 $ & even & $ 27106.1$ & $ 27641.0$ & $ 24651.4$ & $ 10.0$ & $ 12.1$ \\
 $ 12$ & $\mathrm{3p^6\,3d^7}$              & $\mathrm{^2D_a}$ & $5/2 $ & even & $ 27855.5$ & $ 28316.3$ & $ 27096.5$ & $  2.8$ & $  4.5$ \\
 $ 13$ & $\mathrm{3p^6\,3d^7}$              & $\mathrm{^2D_a}$ & $3/2 $ & even & $ 29754.2$ & $ 30453.3$ & $ 28777.7$ & $  3.4$ & $  5.8$ \\
 $ 14$ & $\mathrm{3p^6\,3d^7}$              & $\mathrm{^2H}$   & $11/2$ & even & $ 29856.6$ & $ 30654.6$ & $ 26649.1$ & $ 12.0$ & $ 15.0$ \\
 $ 15$ & $\mathrm{3p^6\,3d^7}$              & $\mathrm{^2H}$   & $9/2 $ & even & $ 30766.3$ & $ 31664.1$ & $ 27677.6$ & $ 11.2$ & $ 14.4$ \\
 $ 16$ & $\mathrm{3p^6\,3d^7}$              & $\mathrm{^2F}$   & $5/2 $ & even & $ 46822.8$ & $ 48435.5$ & $ 43437.5$ & $  7.8$ & $ 11.5$ \\
 $ 17$ & $\mathrm{3p^6\,3d^7}$              & $\mathrm{^2F}$   & $7/2 $ & even & $ 47307.1$ & $ 48984.3$ & $ 43858.6$ & $  7.9$ & $ 11.7$ \\
 $ 18$ & $\mathrm{3p^6\,3d^7}$              & $\mathrm{^2D_b}$ & $3/2 $ & even & $ 69463.3$ & $ 71680.0$ & $ 67360  $ & $  3.1$ & $  6.4$ \\
 $ 19$ & $\mathrm{3p^6\,3d^7}$              & $\mathrm{^2D_b}$ & $5/2 $ & even & $ 70208.8$ & $ 72545.1$ & $ 67989.8$ & $  3.3$ & $  6.7$ \\
 $ 20$ & $\mathrm{3p^6\,3d^6\,(^5D)\,4s}$   & $\mathrm{^6D}$   & $9/2 $ & even & $104016.8$ & $113059.3$ & $110410.6$ & $ -5.8$ & $  2.4$ \\
 $ 21$ & $\mathrm{3p^6\,3d^6\,(^5D)\,4s}$   & $\mathrm{^6D}$   & $7/2 $ & even & $104714.6$ & $113865.8$ & $111195.8$ & $ -5.8$ & $  2.4$ \\
 $ 22$ & $\mathrm{3p^6\,3d^6\,(^5D)\,4s}$   & $\mathrm{^6D}$   & $5/2 $ & even & $105229.8$ & $114458.9$ & $111763.3$ & $ -5.8$ & $  2.4$ \\
 $ 23$ & $\mathrm{3p^6\,3d^6\,(^5D)\,4s}$   & $\mathrm{^6D}$   & $3/2 $ & even & $105586.0$ & $114868.0$ & $112151.9$ & $ -5.9$ & $  2.4$ \\
 $ 24$ & $\mathrm{3p^6\,3d^6\,(^5D)\,4s}$   & $\mathrm{^6D}$   & $1/2 $ & even & $105795.5$ & $115108.6$ & $112379.3$ & $ -5.9$ & $  2.4$ \\
 $ 25$ & $\mathrm{3p^6\,3d^6\,(^5D)\,4s}$   & $\mathrm{^4D}$   & $7/2 $ & even & $116491.5$ & $125023.6$ & $120909.5$ & $ -3.7$ & $  3.4$ \\
 $ 26$ & $\mathrm{3p^6\,3d^6\,(^5D)\,4s}$   & $\mathrm{^4D}$   & $5/2 $ & even & $117298.8$ & $125960.5$ & $121807.7$ & $ -3.7$ & $  3.4$ \\
 $ 27$ & $\mathrm{3p^6\,3d^6\,(^5D)\,4s}$   & $\mathrm{^4D}$   & $3/2 $ & even & $117832.2$ & $126574.9$ & $122386.1$ & $ -3.7$ & $  3.4$ \\
 $ 28$ & $\mathrm{3p^6\,3d^6\,(^5D)\,4s}$   & $\mathrm{^4D}$   & $1/2 $ & even & $118139.8$ & $126928.8$ & $122717.4$ & $ -3.7$ & $  3.4$ \\
 $ 29$ & $\mathrm{3p^6\,3d^6\,(^3P_a)\,4s}$ & $\mathrm{^4P}$   & $5/2 $ & even & $135512.9$ & $144723.7$ & $139289.4$ & $ -2.7$ & $  3.9$ \\
 $ 30$ & $\mathrm{3p^6\,3d^6\,(^3P_a)\,4s}$ & $\mathrm{^4P}$   & $3/2 $ & even & $135781.1$ & $145030.5$ & $139619.2$ & $ -2.7$ & $  3.9$ \\
 $ 31$ & $\mathrm{3p^6\,3d^6\,(^3H)\,4s}$   & $\mathrm{^4H}$   & $13/2$ & even & $136008.9$ & $145290.5$ & $139886.7$ & $ -2.8$ & $  3.9$ \\
 $ 32$ & $\mathrm{3p^6\,3d^6\,(^3P_a)\,4s}$ & $\mathrm{^4P}$   & $1/2 $ & even & $136201.8$ & $145512.5$ & $140140.9$ & $ -2.8$ & $  3.8$ \\
 $ 33$ & $\mathrm{3p^6\,3d^6\,(^3H)\,4s}$   & $\mathrm{^4H}$   & $11/2$ & even & $138011.2$ & $147167.4$ & $138446.2$ & $ -0.3$ & $  6.3$ \\
 $ 34$ & $\mathrm{3p^6\,3d^6\,(^3F_a)\,4s}$ & $\mathrm{^4F}$   & $9/2 $ & even & $139513.2$ & $148730.2$ & $141220.3$ & $ -1.2$ & $  5.3$ \\
 $ 35$ & $\mathrm{3p^6\,3d^6\,(^3H)\,4s}$   & $\mathrm{^4H}$   & $7/2 $ & even & $139826.4$ & $149097.6$ & $141577.2$ & $ -1.2$ & $  5.3$ \\
 $ 36$ & $\mathrm{3p^6\,3d^6\,(^3H)\,4s}$   & $\mathrm{^4H}$   & $9/2 $ & even & $139836.5$ & $149191.9$ & $140343  $ & $ -0.4$ & $  6.3$ \\
 $ 37$ & $\mathrm{3p^6\,3d^6\,(^3F_a)\,4s}$ & $\mathrm{^4F}$   & $7/2 $ & even & $140084.8$ & $149395.2$ & $141832  $ & $ -1.2$ & $  5.3$ \\
 $ 38$ & $\mathrm{3p^6\,3d^6\,(^3F_a)\,4s}$ & $\mathrm{^4F}$   & $5/2 $ & even & $140286.6$ & $149629.3$ & $142023.5$ & $ -1.2$ & $  5.4$ \\
 $ 39$ & $\mathrm{3p^6\,3d^6\,(^3F_a)\,4s}$ & $\mathrm{^4F}$   & $3/2 $ & even & $140979.3$ & $150489.1$ & $141561.2$ & $ -0.4$ & $  6.3$ \\
 $ 40$ & $\mathrm{3p^6\,3d^6\,(^3P_b)\,4s}$ & $\mathrm{^2P}$   & $3/2 $ & even & $142408.9$ & $151421.3$ & $144815.1$ & $ -1.7$ & $  4.6$ \\
 $ 41$ & $\mathrm{3p^6\,3d^6\,(^3G)\,4s}$   & $\mathrm{^4G}$   & $11/2$ & even & $143189.5$ & $152259.5$ & $145702.2$ & $ -1.7$ & $  4.5$ \\
 $ 42$ & $\mathrm{3p^6\,3d^6\,(^3G)\,4s}$   & $\mathrm{^4G}$   & $9/2 $ & even & $143424.1$ & $152637.1$ & $145962.5$ & $ -1.7$ & $  4.6$ \\
 $ 43$ & $\mathrm{3p^6\,3d^6\,(^3G)\,4s}$   & $\mathrm{^4G}$   & $7/2 $ & even & $143641.9$ & $152973.0$ & $146194.3$ & $ -1.7$ & $  4.6$ \\
 $ 44$ & $\mathrm{3p^6\,3d^6\,(^3G)\,4s}$   & $\mathrm{^4G}$   & $5/2 $ & even & $143766.9$ & $153187.4$ & $146061.5$ & $ -1.6$ & $  4.9$ \\
 $ 45$ & $\mathrm{3p^6\,3d^6\,(^3H)\,4s}$   & $\mathrm{^2P}$   & $1/2 $ & even & $143846.5$ & $153267.4$ & $146153.8$ & $ -1.6$ & $  4.9$ \\
 $ 46$ & $\mathrm{3p^6\,3d^6\,(^3H)\,4s}$   & $\mathrm{^2H}$   & $11/2$ & even & $145813.1$ & $154764.1$ & $145192.1$ & $  0.4$ & $  6.6$ \\
 $ 47$ & $\mathrm{3p^6\,3d^6\,(^3H)\,4s}$   & $\mathrm{^2H}$   & $9/2 $ & even & $147063.6$ & $156007.6$ & $147635.9$ & $ -0.4$ & $  5.7$ \\
 $ 48$ & $\mathrm{3p^6\,3d^6\,(^3F_b)\,4s}$ & $\mathrm{^2F}$   & $7/2 $ & even & $147768.3$ & $156833.6$ & $148358.2$ & $ -0.4$ & $  5.7$ \\
 $ 49$ & $\mathrm{3p^6\,3d^6\,(^3F_b)\,4s}$ & $\mathrm{^2F}$   & $5/2 $ & even & $147948.1$ & $157127.4$ & $        $ & $  -  $ & $   - $ \\
 $ 50$ & $\mathrm{3p^6\,3d^6\,(^3D)\,4s}$   & $\mathrm{^4D}$   & $3/2 $ & even & $150422.2$ & $159451.6$ & $151574.7$ & $ -0.8$ & $  5.2$ \\
 \hline
\end{tabular}

\flushleft{Key: $i$: level index;
Conf: dominant electron configuration;
Term: dominant LS term;
J: level angular momentum;
GRASP: present GRASP calculation, 
AS: present {\sc autostructure} calculation; 
NIST: recommended value from NIST data base~\citep{nist2018};
$\%$: deviation respect the recommended values of NIST, in percentage.
All energies in $\mathrm{cm}^{-1}$.
}

\end{table*}

To perform the close-coupling (CC) integration including all the $6\,841$
levels obtained in the atomic structure is beyond the capabilities of existing 
workstations and even supercomputers.
Consequently, we have selected the lowest excited $262$ levels for the CC expansion.
For analysis that favours the ground state and first few metastable states the 
completeness of this CC expansion is acceptable.

The online material will present a table of oscillator $f$ strengths and 
Einstein spontaneous emission coefficients $A$ for all the transitions between 
the $262$ lowest-excited levels.
We show the values obtained with both methods {\sc grasp} and {\sc autostructure}.
This comparison gives an idea of the consistency for energies and transition 
probabilities for both atomic models.
We also compare our results for the Einstein A-coefficients with previous 
theoretical calculations in the literature from \citet{hansen1984}.
Unfortunately, to the best our knowledge there are no experimental data available 
in the scientific literature for $\mathrm{Ni}^{3+}$ to compare with.

Finally, to perform the scattering calculation we have shifted our calculated
energies for the levels included in the CC expansion to the observed values of the 
NIST data base.
Doing so we make sure that the calculated wave lengths for the transitions will
fit exactly with the observed ones,
which is the requested for proper modelling of the astrophysical objects.
In the NIST database, $\mathrm{Ni^{3+}}$ has some missing energy levels for 
the highly excited states.
Therefore, in those cases we have shifted our theoretical values by the 
difference with respect to the known NIST levels.
We compare our final results using the shifted target energies with the unshifted
ones as a test of accuracy.

\section{Scattering and photoionization processes}
\label{sec:scattering}

We use an $R$-matrix formalism \citep{hummer1993,berrington1995}.
In the inner region, we use the fully relativistic {\sc darc} code 
\citep{ait-tahar1996,norrington1981,norrington1987}
to get the stationary solutions of the $N+1$ electron atom.
We calculate the $N+1$ wave functions by diagonalization of the $N+1$ electron
Hamiltonian.
In addition, we calculate the dipole momentum matrices for the relevant 
photoionization transitions.
In the outer region, we use the parallel version of the {\sc stgf} program 
to calculate the EIE collision strengths $\Omega$, 
and the radiative damped version {\sc pstgbf0damp} for the photoionization cross sections.

We calculate the photoionization cross sections from several initial states 
of the parent ion $\mathrm{Ni}^{2+}$.
These levels are the relevant ones for an opacity model.
With the available computational resources it is absolutely impossible to include
in the close coupling (CC) expansion all the $6\,841$ levels calculated with the
previous described CI expansion in {\sc grasp}.
To have a reasonable accuracy in the calculation compatible with an affordable 
computation cost we have selected the $262$ levels with the lowest energy for
the CC expansion.

We use the same set of $262$ levels to calculate the electron-impact
excitation of $\mathrm{Ni}^{3+}$.
We include partial waves with angular momentum up to $J=36$.

\subsection{Inner region}
\label{subsec:inner}

For our {\sc darc} calculation the R-Matrix inner-region radius is set to $59.52\au$.
We calculate the Hamiltonian matrices and the transition dipole momentum matrices.
Including the first 262 levels of $\mathrm{Ni}^{3+}$ target in the CC expansion
we get a maximum of $1\,818$ channels in each $J^{\pi}$ symmetry.

For the photoionization calculation we calculate partial waves with a 
total angular momentum of $J=0-5$ and both parities.
The lowest levels of the $\mathrm{Ni}^{2+}$ ion have an angular momentum of
$J=0-4$ and even parity (see \citet{nist2018}).
Levels with higher angular momenta are very excited and they will 
rapidly decay to lower $J$ by an M1 or E2 transition.
Levels with odd parity are very high in energy,
the first one is the $\mathrm{3p^6\,3d^7\,4p\,^5F_5^{o}}$ with an
energy of $1.0043 \Ry$ relative to the ground state.
They will be connected by an E1 transition to any lower level with 
even parity and their population will be zero in any astrophysical object.
In addition, for each partial wave we calculate the dipole matrices
with all their possible E1 couples. 

For the electron-impact excitation calculations we need a more extended set of 
partial waves. 
We have calculated the energies and wave functions of the channels of all partial
waves with an angular momentum of $J=0-36$ and both parities plus a top-up.

\subsection{Outer region}
\label{subsec:outer}

To calculate the photoionization cross sections as a function of photon energy 
in Rydbergs we utilise the parallel version of {\sc pstgbf0-damp},
a code which calculates the photoionization cross sections
utilizing the previously calculated bound-free matrix elements.
The first serial version of {\sc stgbf0damp} was by 
Gorczyca and Badnell (unpublished).
The first stage is to determine the bound levels of the ($N+1$)-electron system 
$\mathrm{Ni}^{2+}$ in the program {\sc stgb} \citep{seaton1982,berrington1987}, 
which reads the wave functions for 
a specific partial wave in the inner region and determines its bound states.
In our final calculated cross sections, we shift the energies of the numerical 
$\mathrm{Ni}^{2+}$ levels to fit exactly the ionization potential with the
values tabulated in NIST data basis.
Hence the threshold of the cross sections fit exactly with the ionization
potential of the initial state.

We split the energy range into two regions. 
In the low-energy region we adopt a fine energy mesh 
of $1.5 \times 10^{-5}\,z^2 \Ry$, being $z=3$ the charge of the final ion,
to properly resolve the resonance structures 
converging onto the target thresholds. 
A linear grid with a total of $40\,000$ energy points was included up to the 
excitation energy of the last level included in the close-coupling expansion. 
Above this threshold resonances are not present and the cross sections are smoother,
hence a coarser mesh of $3 \times 10^{-3}\,z^2 \Ry$ was utilised.
For higher photon energies, above the excitation of the last included level 
in our CC expansion, $5.5 \Ry$ in our case,
there are more possible processes present in nature,
for example the ionization with a final level which is not included in 
our CC expansion, or double ionization.
Higher excited states, for example excitations $\mathrm{3s}^{-1}$,
while included in the CI expansion, are not included in the CC.
Due to the limitation of our CC expansion, these processes can not be 
reproduced by our model.
Hence present results are valid for a maximum photon 
energy of $5.5 \Ry$, approximately twice the ionization energy 
of $\mathrm{Ni}^{2+}$ from its ground level.

For the electron-impact excitation evaluation the parallel version of the 
{\sc stgf} undamped package 
(Seaton {\it et al.}, unpublished)
was utilised in the outer region.
{\sc pstgf} calculates the outer-region wave function using a Numerov method
and including the coupling in the outer-region as a perturbation.
{\sc pstgf} joins the calculated wave function with one in the inner region 
in terms of the $R$-Matrix method \citep{burke2011}.
In the outer region problem high angular momenta do not contribute to the 
resonance structures, hence we restrict the fine-mesh calculation to the low 
partial waves with $J=0-20$ and adopt a fine energy mesh 
of $1.5 \times 10^{-5}\,z^2 \Ry$, $z=3$ being the ion charge,
we incorporate a total of $40\,000$ points in the low energy region.
At higher energies, above the threshold energy of the last level included in 
the CC expansion, there is no more resonance structure and  
the cross sections are smooth, so we use a coarser 
mesh of $3 \times 10^{-3}\,z^2 \Ry$.
The higher angular momenta $J=21-36$ do not contribute to the resonance 
structure, even for low energies,
hence the coarse mesh listed above is sufficient in the whole energy range. 
Finally, to include the remaining angular momenta up to $J$ infinity 
we perform a top-up procedure. 
For dipole allowed transitions we use the Burgess sum rule \cite{burgess1974}
and for the non-dipole allowed transitions with non-zero infinite energy Born limit 
a geometric series~\cite{badnell2001a}.

As the selected CC expansion in the target is considerably smaller than the initial
CI expansion, we expect pseudoresonances to appear for electron final energies 
larger than the energy of the last level included in the CC expansion ($2.57 \Ry$), 
equivalent to an electron temperature of $8 \times 10^{5} \kelvin$.
As the peak-abundance temperature of $\mathrm{Ni}^{3+}$ in a collisional plasma
is $4 \times 10^{4} \kelvin$ \citep{bryans2006}, these pseudoresonances will not
affect the effective collision strengths at temperatures where 
$\mathrm{Ni}^{3+}$ has a significant ionisation fraction.
Nevertheless, we have checked for the relevant transitions, the collision strengths 
these pseudoresonances are present.

\section{Results}
\label{sec:results}

\subsection{Electron-impact excitation of $\mathrm{Ni^{3+}}$}
\label{subsec:Ni3eie}

\begin{figure*}
   \includegraphics[width=0.4\textwidth]{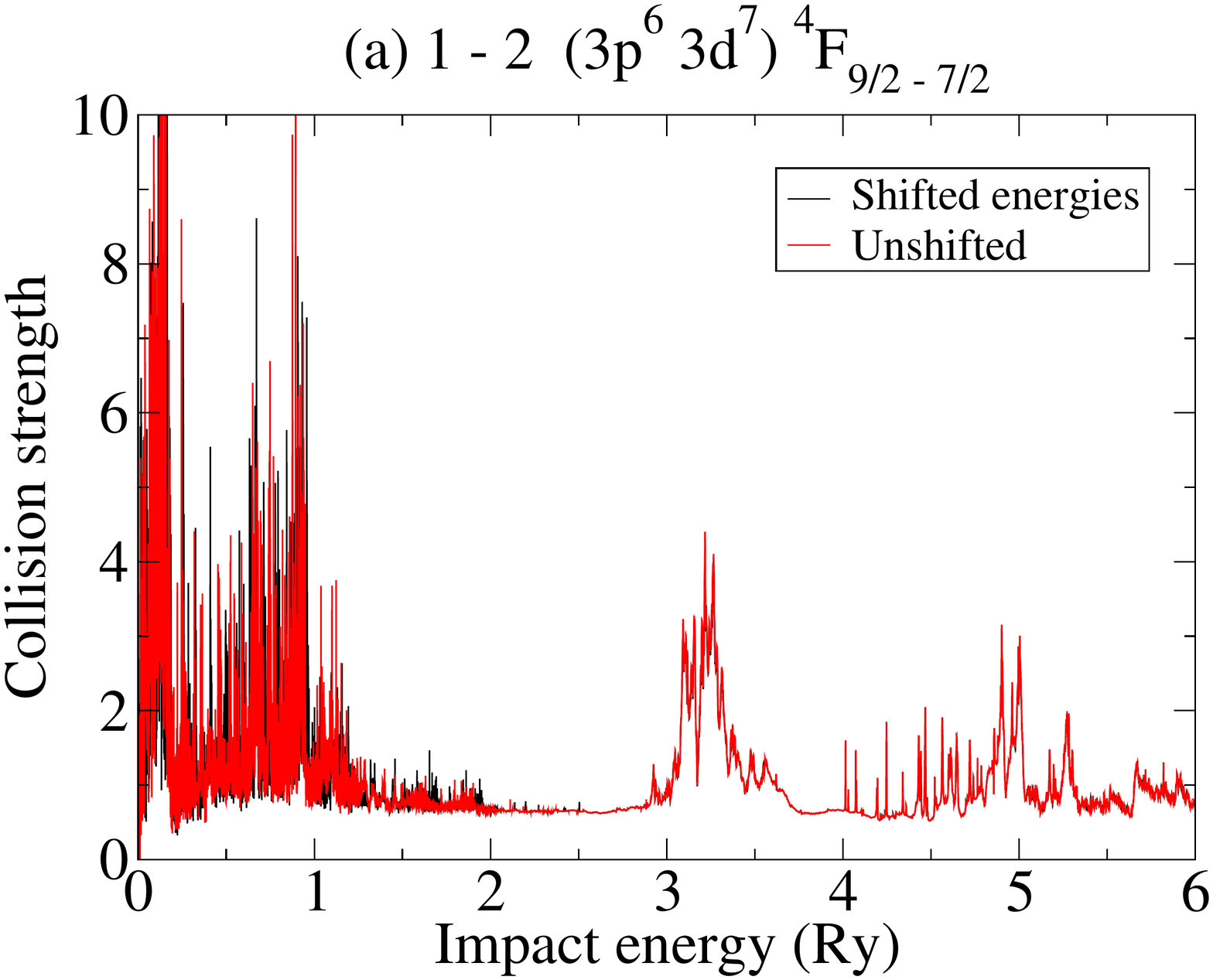} \,
   \includegraphics[width=0.4\textwidth]{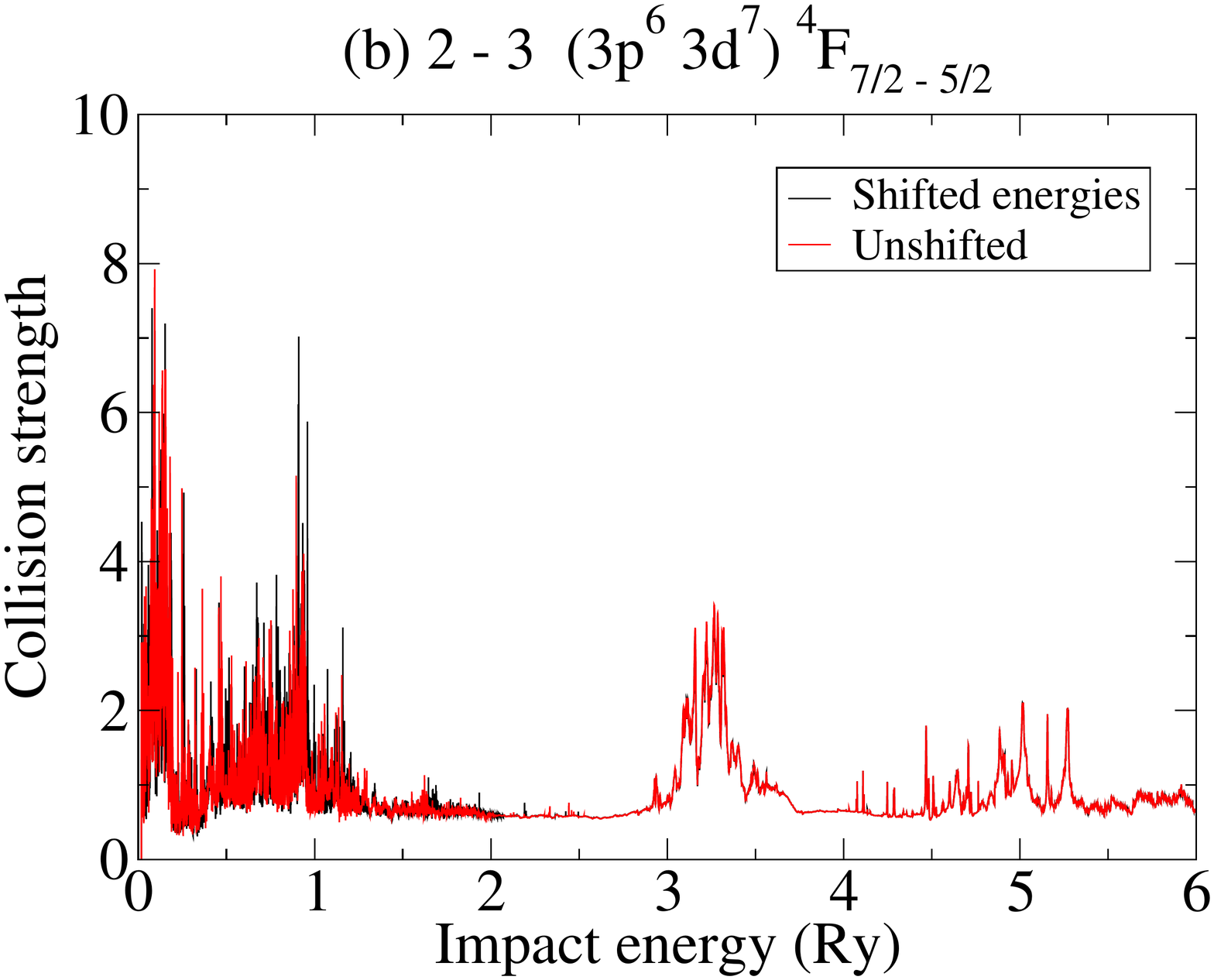} \\
   \includegraphics[width=0.4\textwidth]{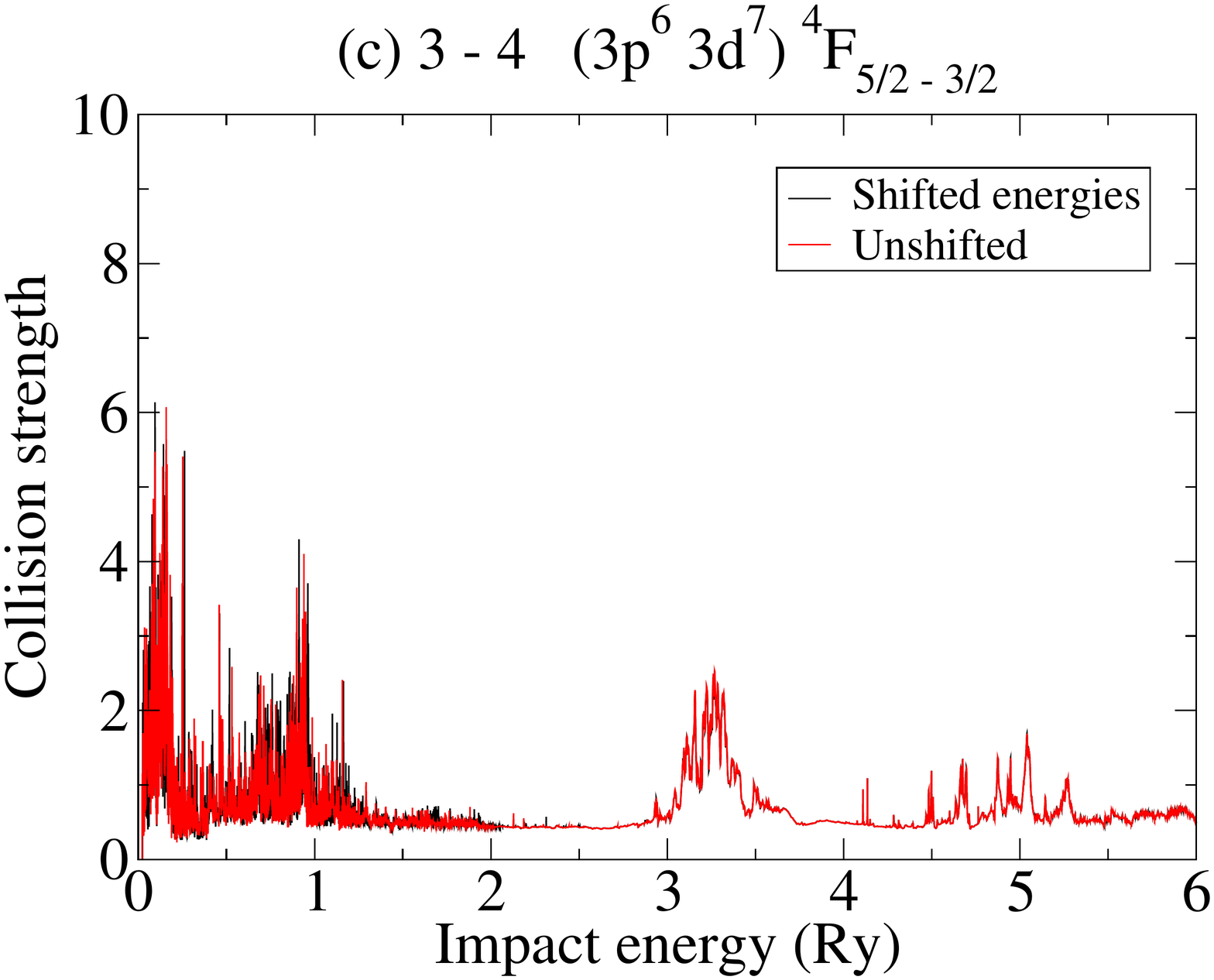} \,
   \includegraphics[width=0.4\textwidth]{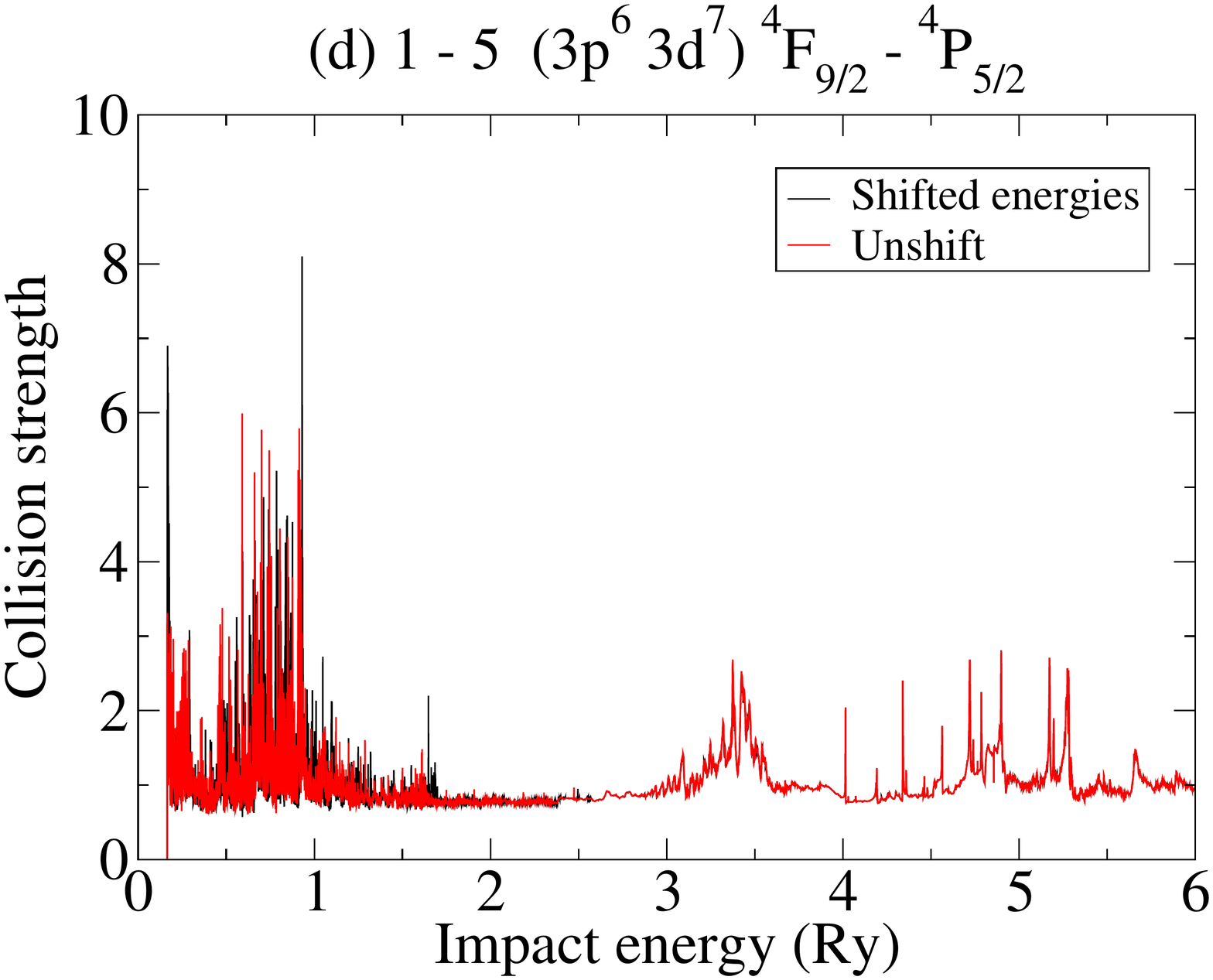} \\
   \includegraphics[width=0.4\textwidth]{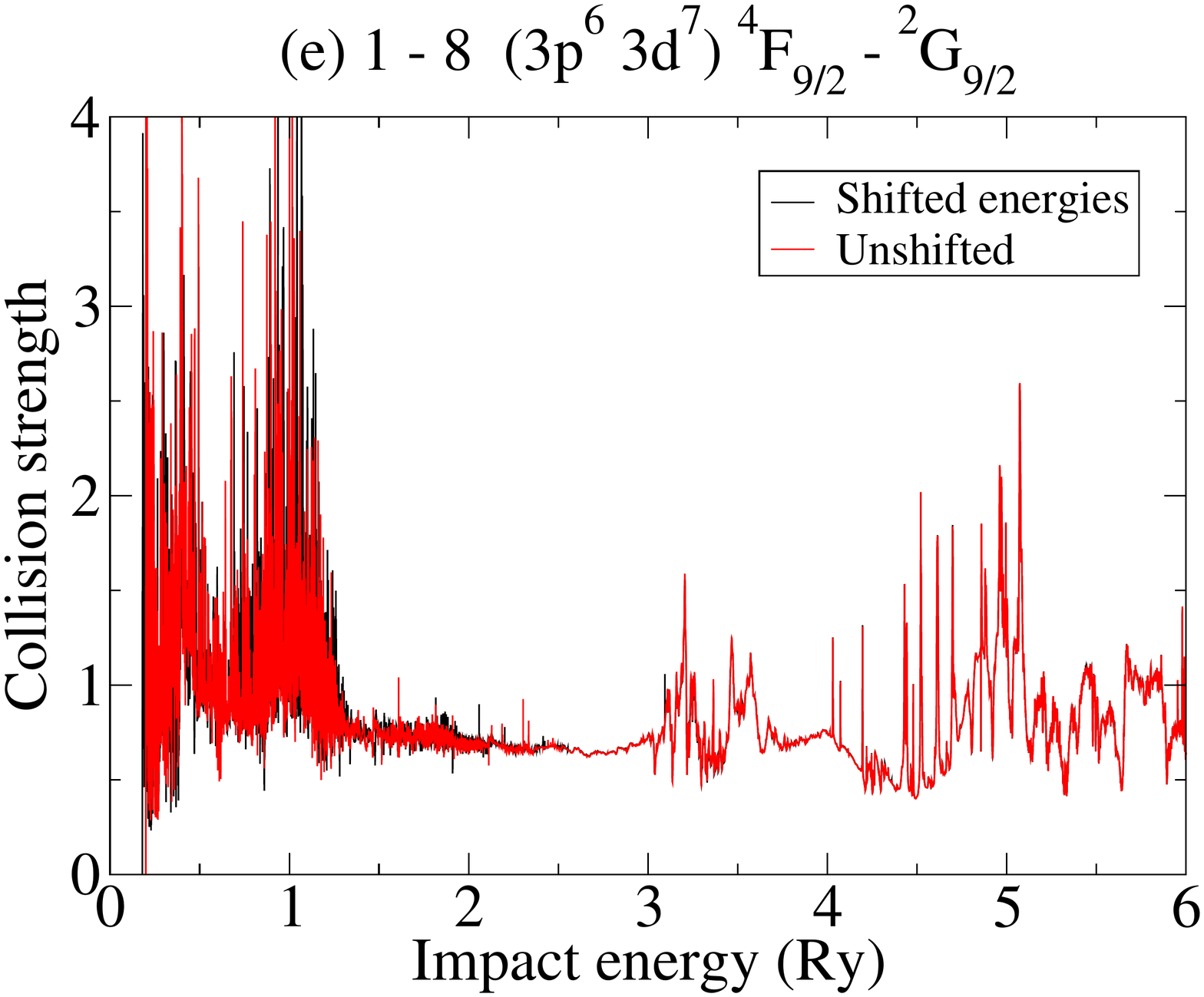} \,
   \includegraphics[width=0.4\textwidth]{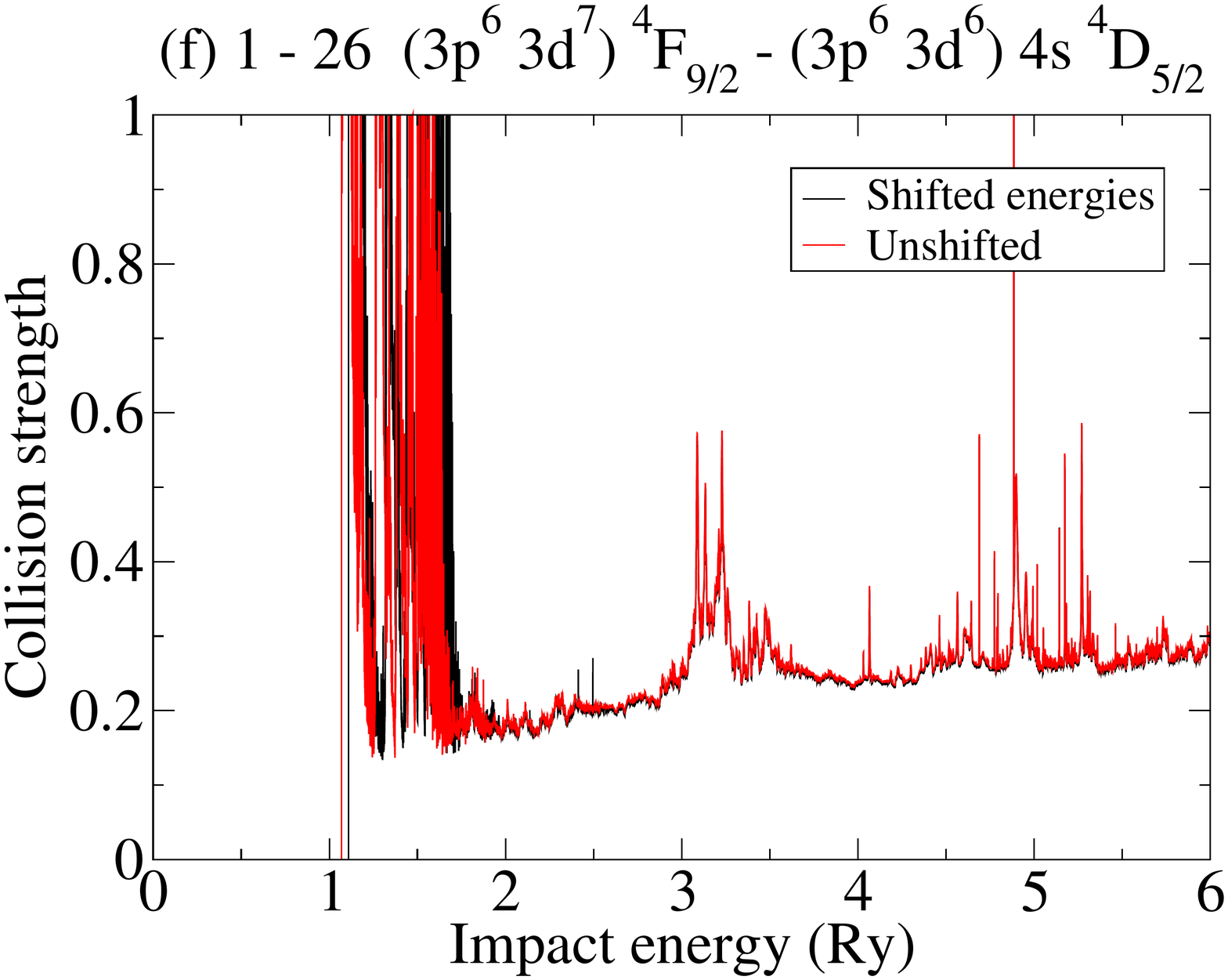} \\
   \includegraphics[width=0.4\textwidth]{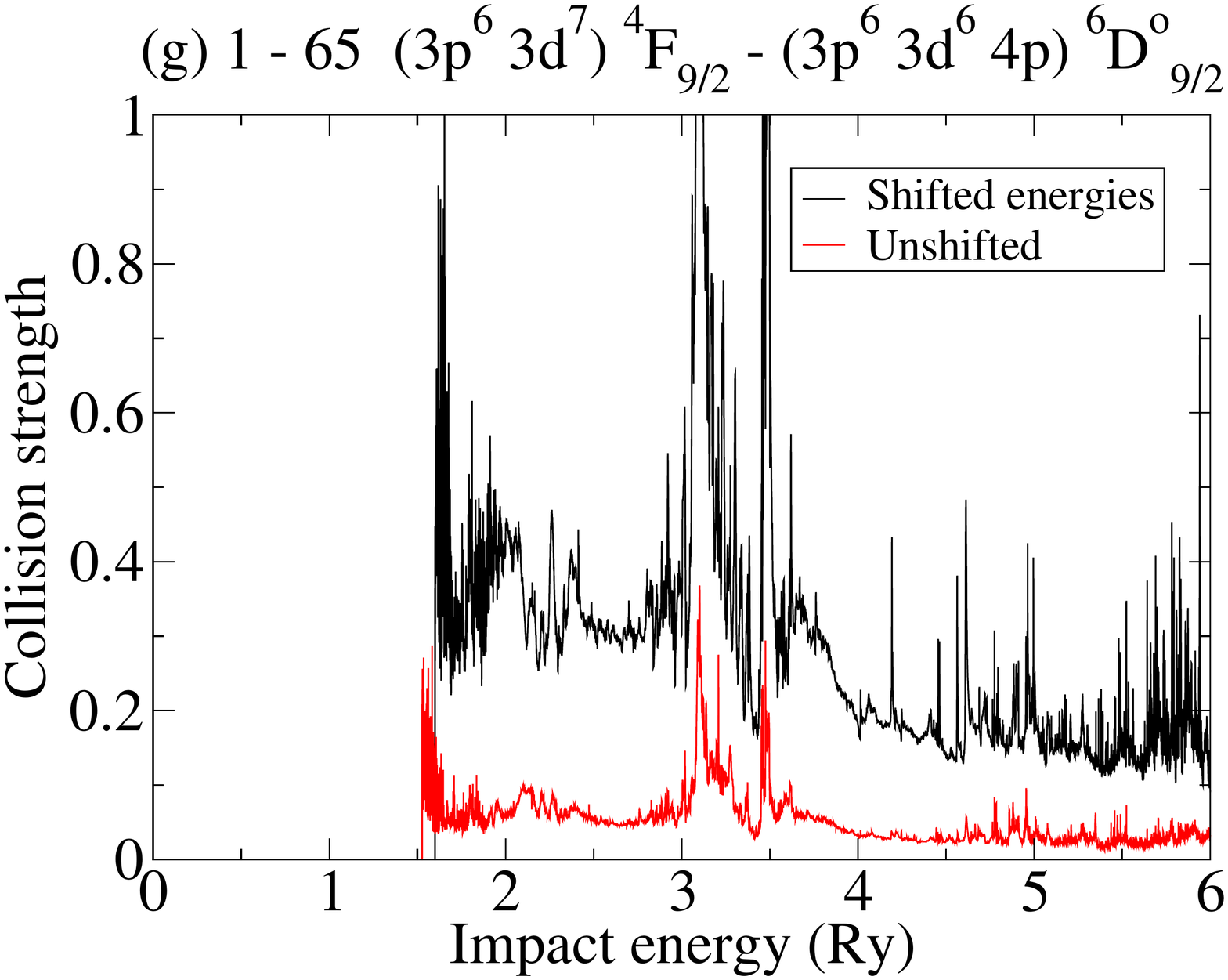} \,
   \includegraphics[width=0.4\textwidth]{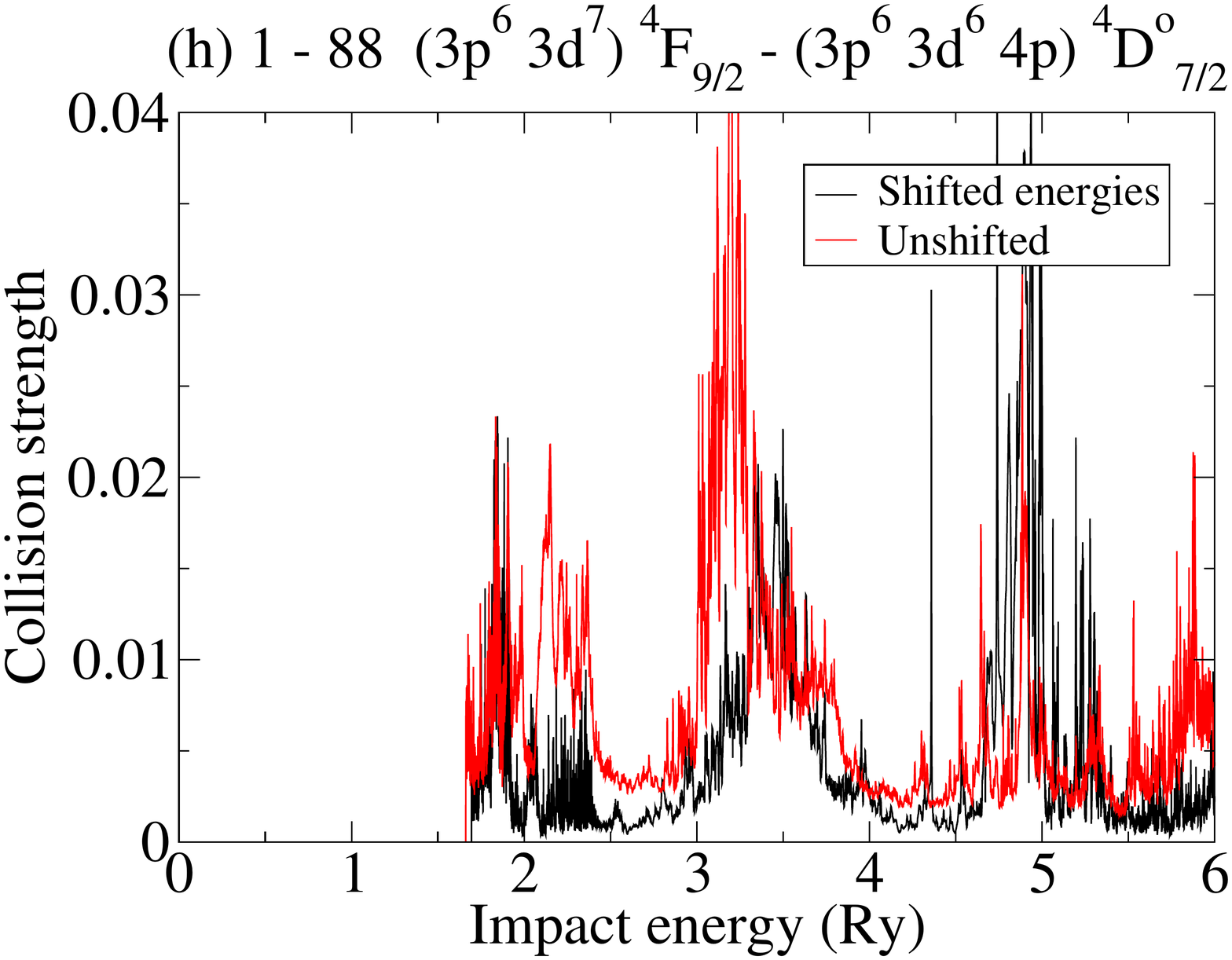} 
\caption{Electron-impact excitation collision strengths $\Omega$
   of $\mathrm{Ni}^{3+}$.}
\label{fig:ni3omg}
\end{figure*}

\begin{figure*}
   \includegraphics[width=0.4\textwidth]{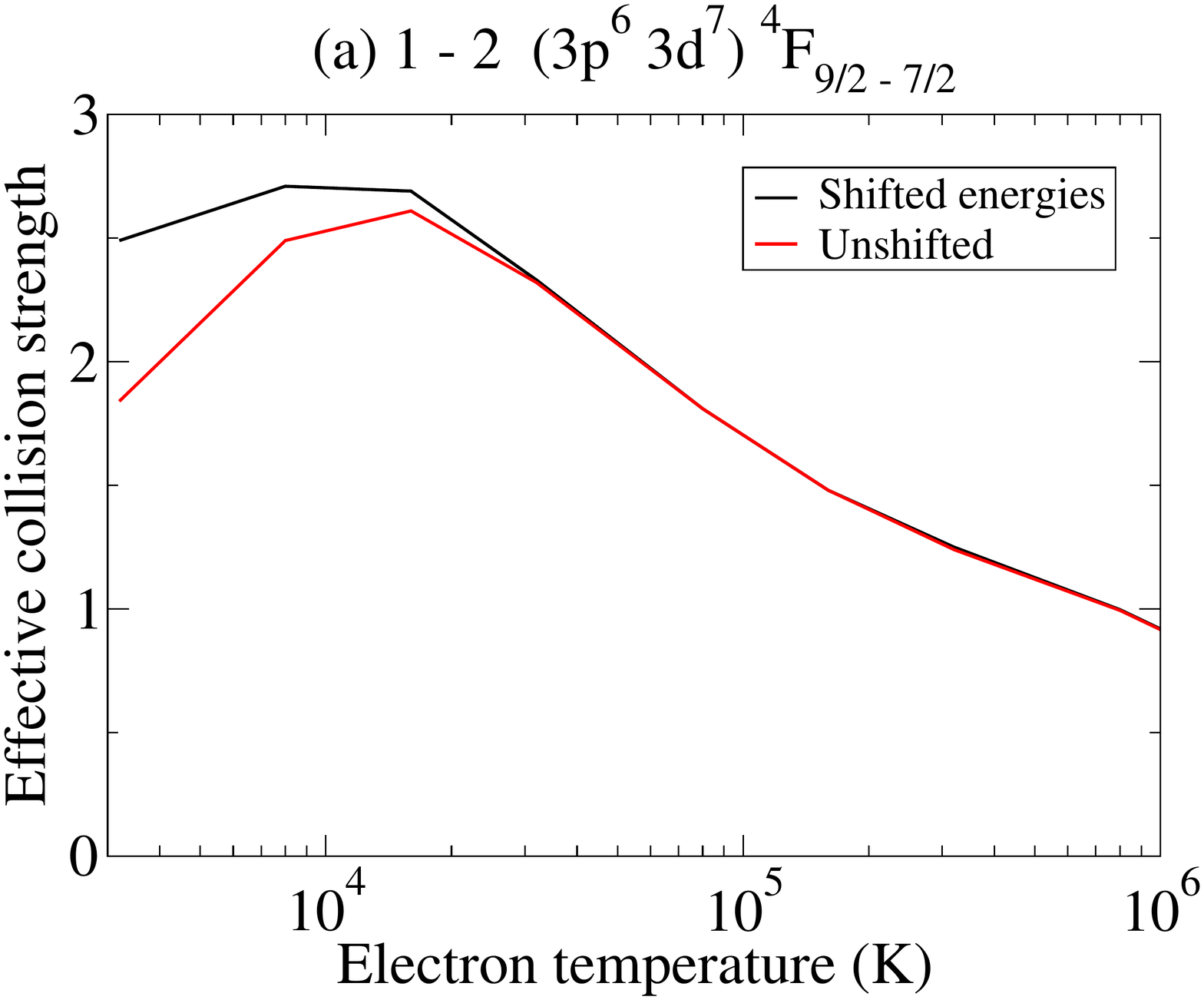} \,
   \includegraphics[width=0.4\textwidth]{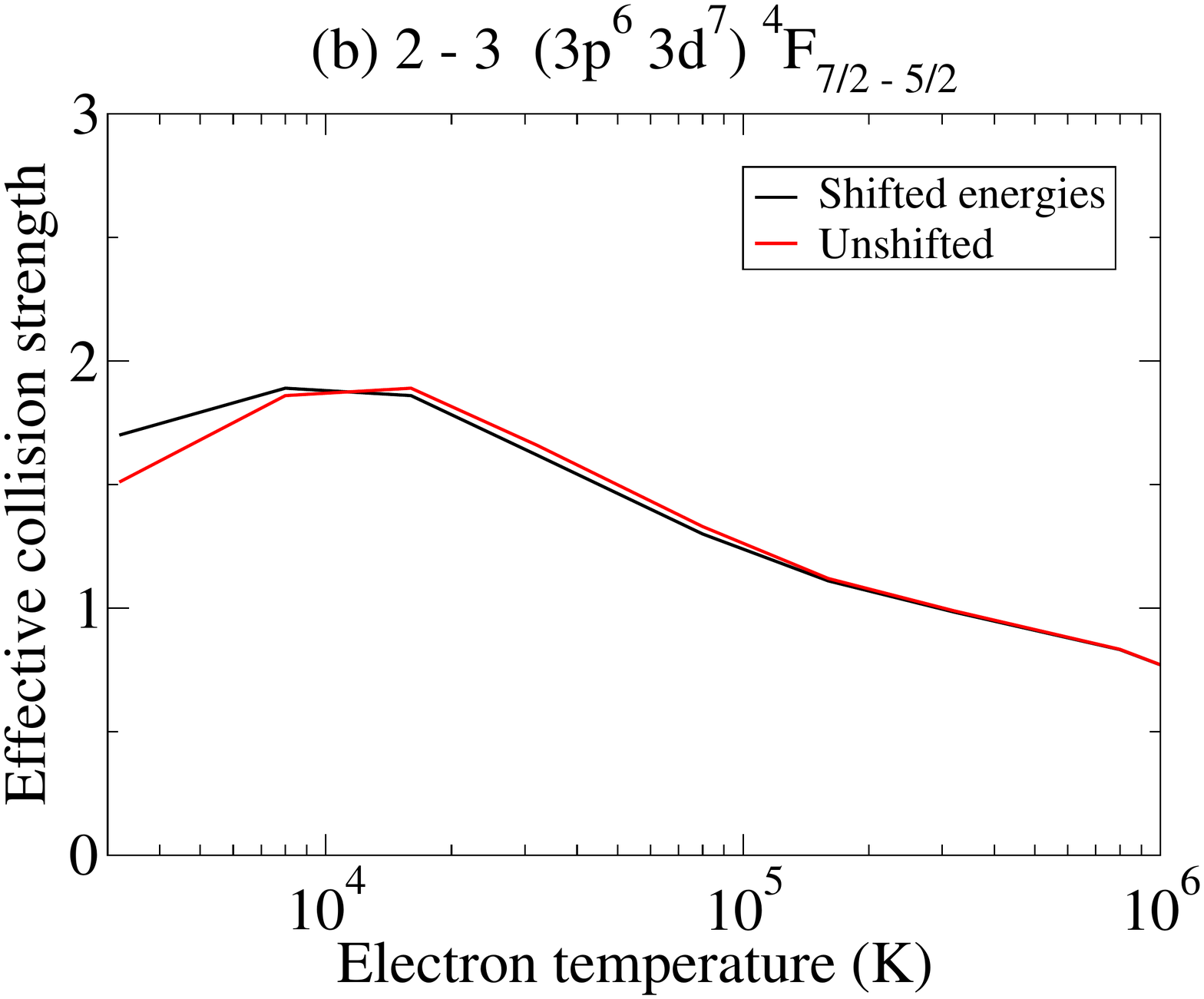} \\
   \includegraphics[width=0.4\textwidth]{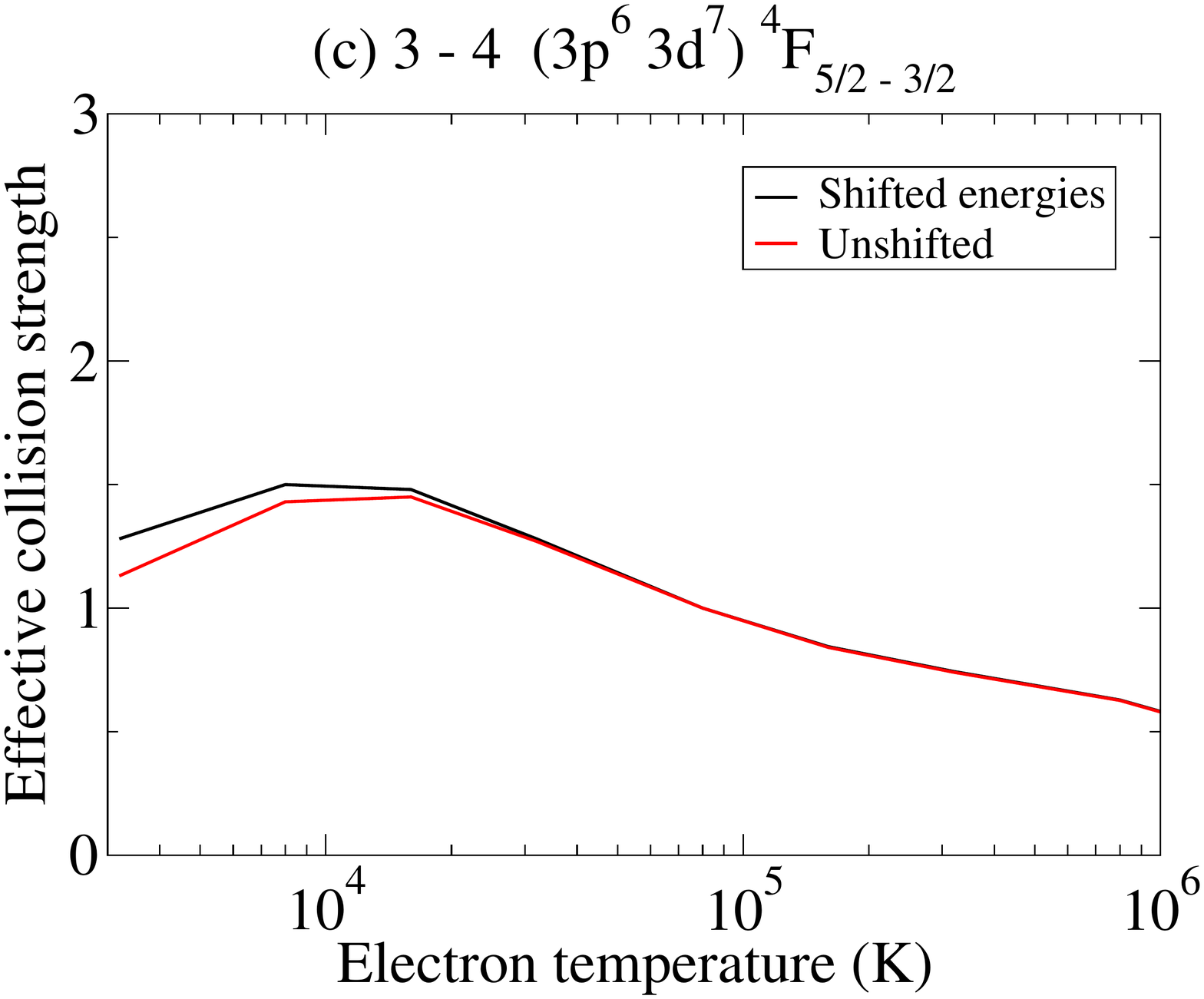} \,
   \includegraphics[width=0.4\textwidth]{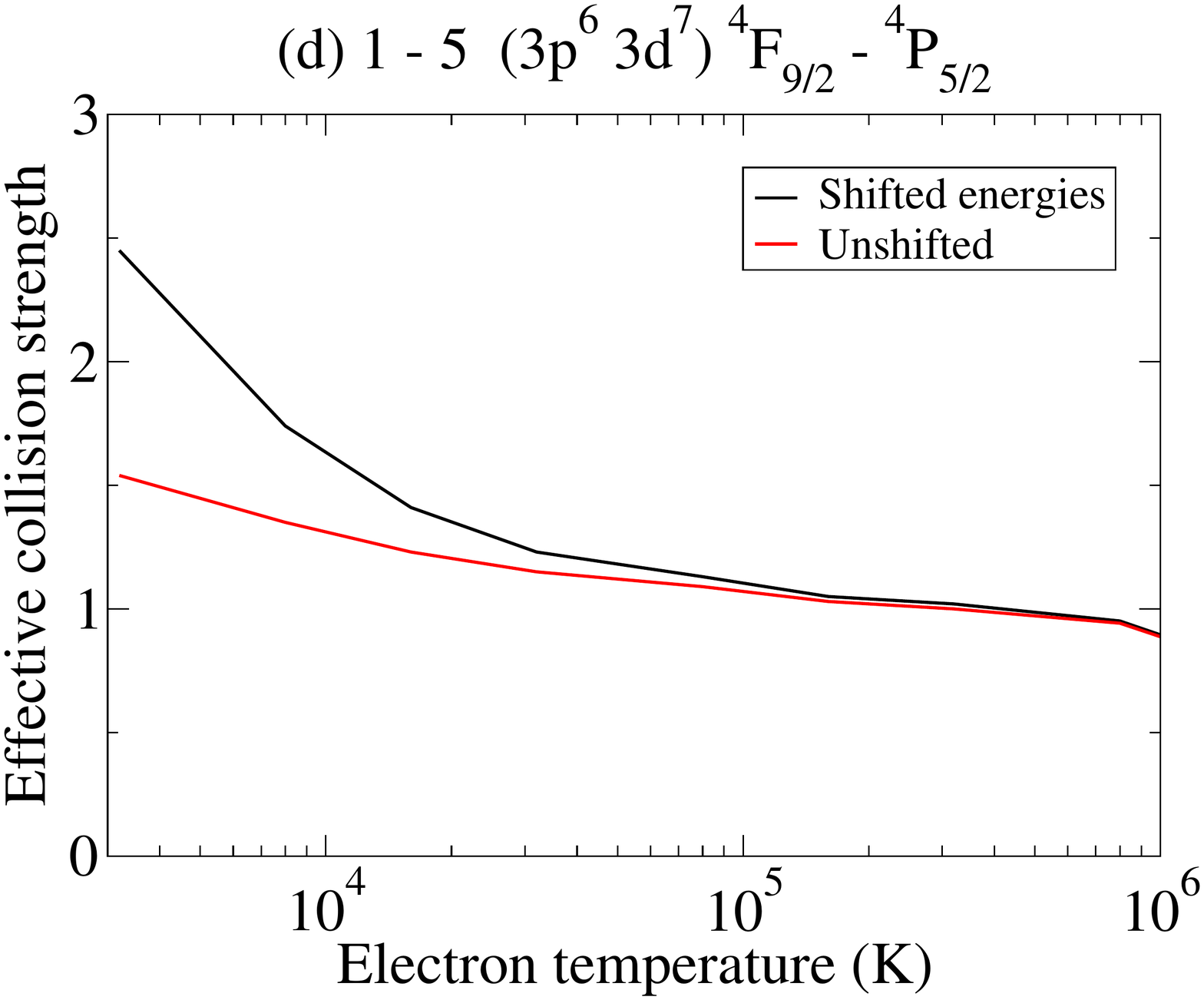} \\
   \includegraphics[width=0.4\textwidth]{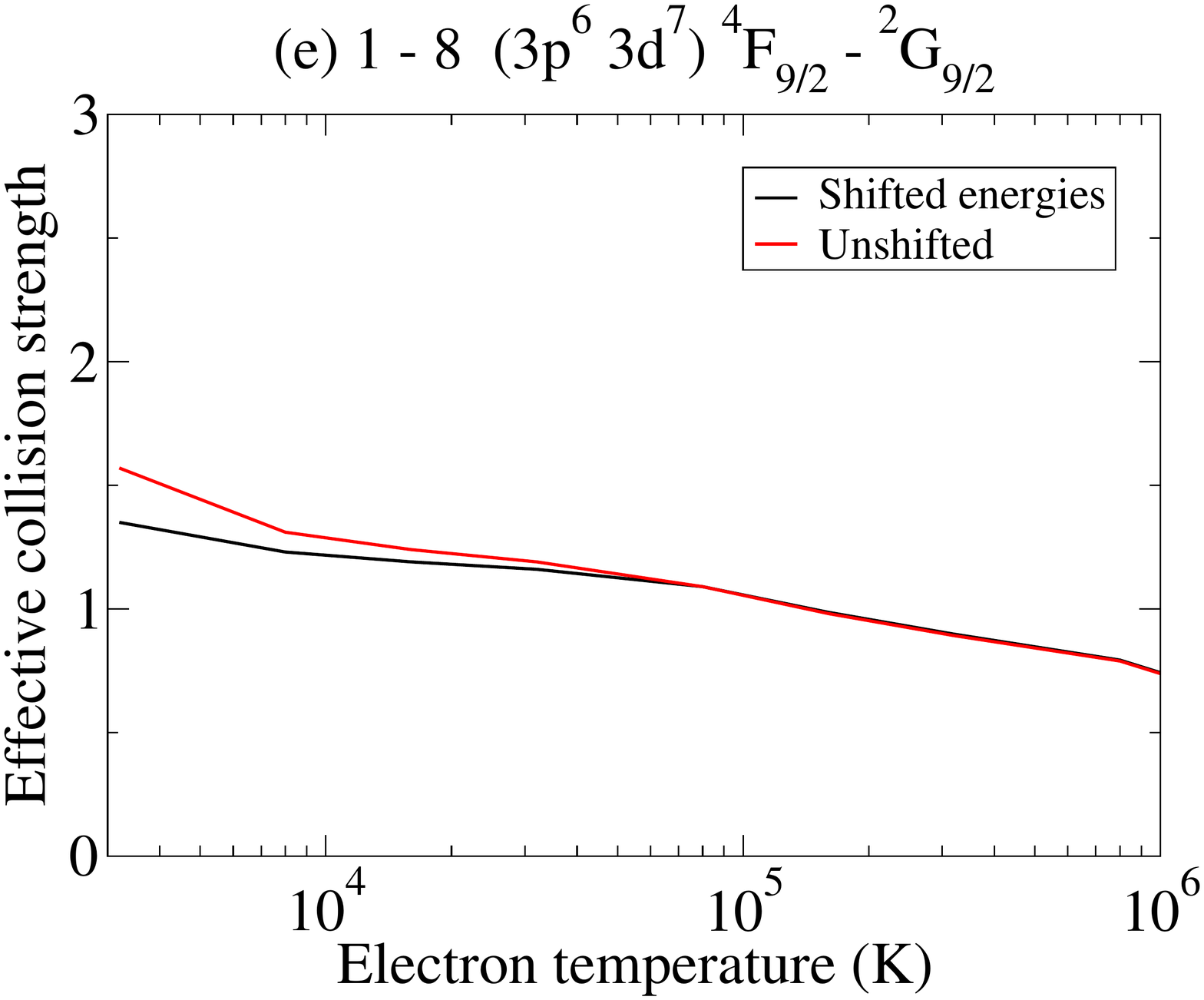} \,
   \includegraphics[width=0.4\textwidth]{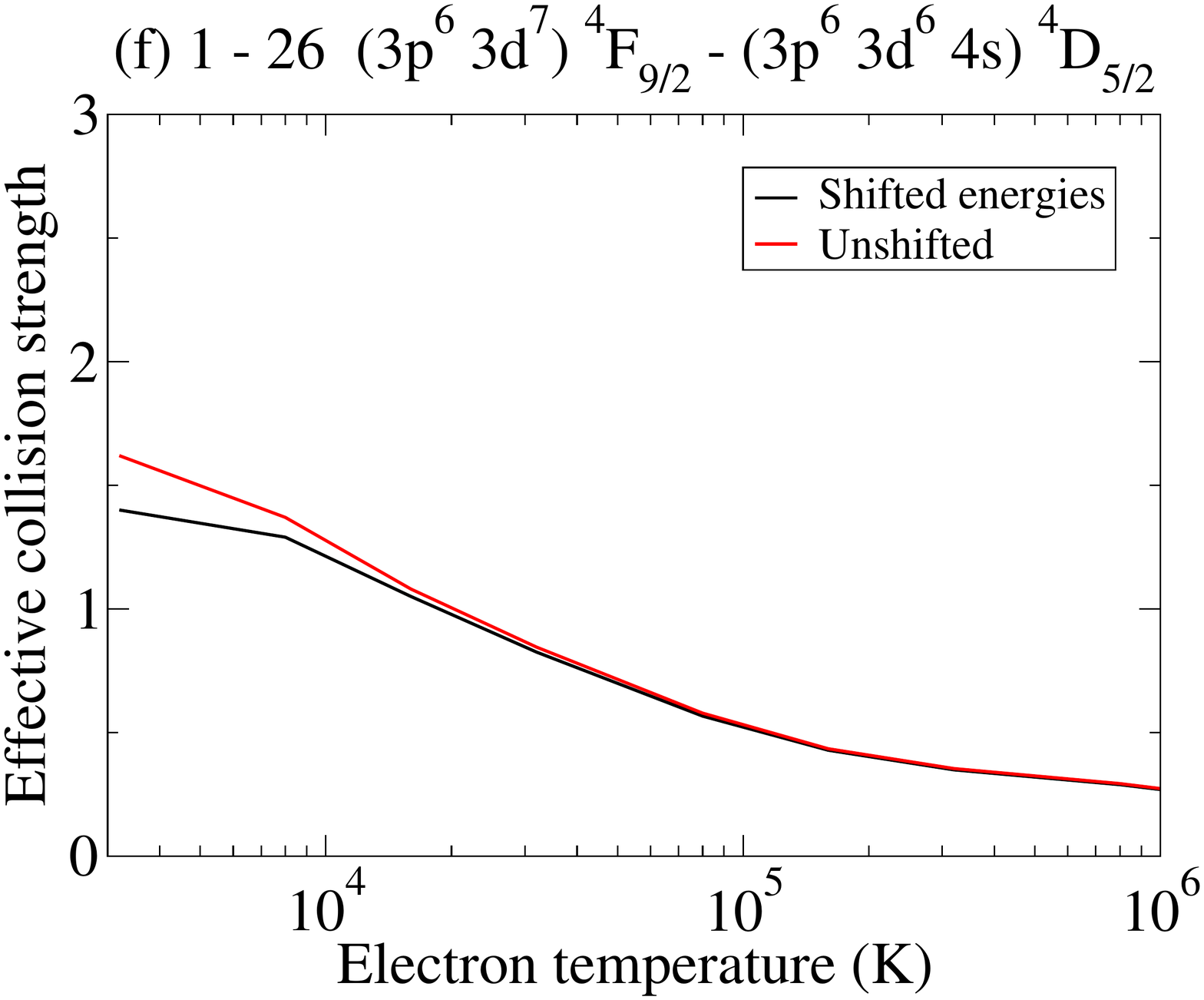} \\
   \includegraphics[width=0.4\textwidth]{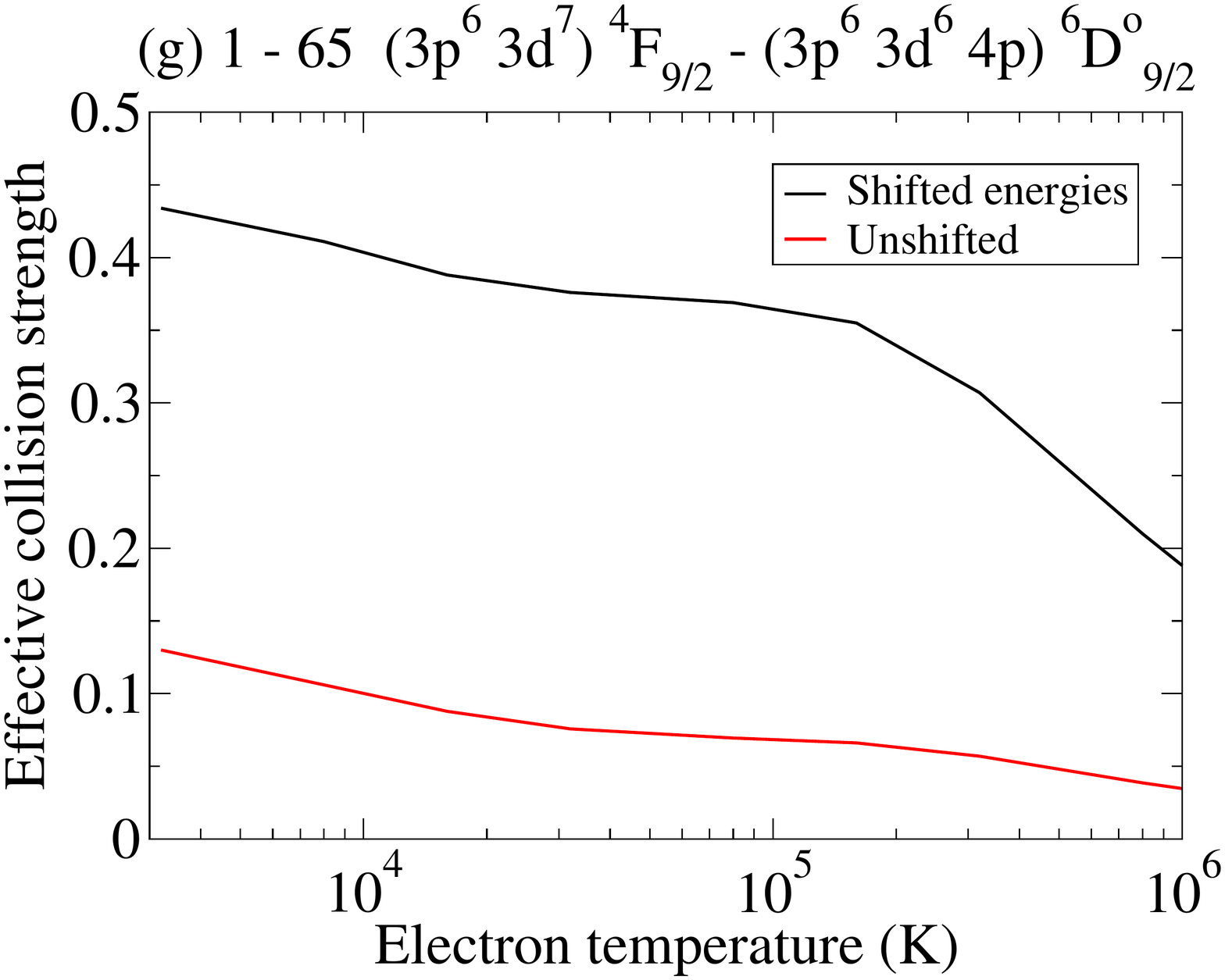} \,
   \includegraphics[width=0.4\textwidth]{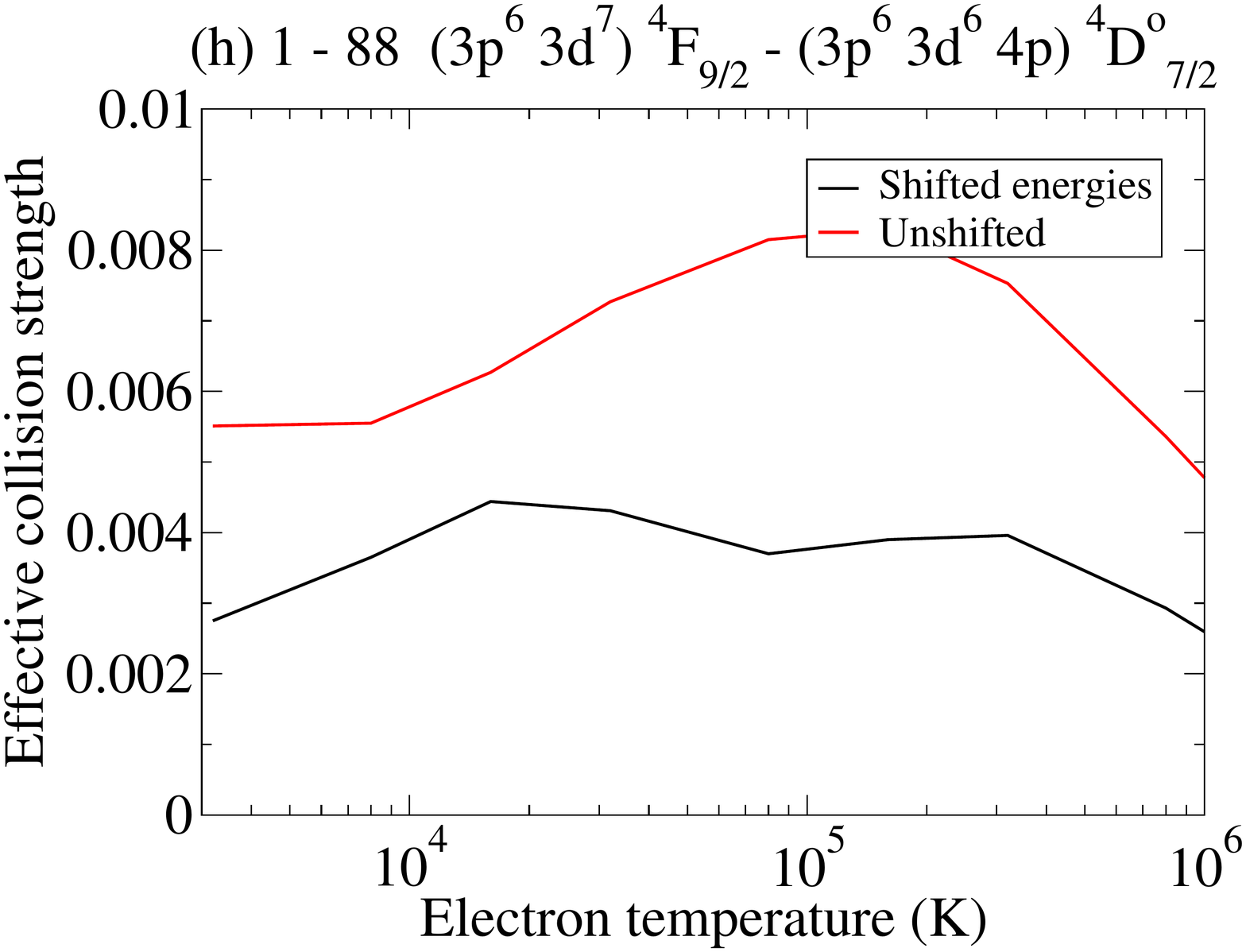} 
\caption{Electron-impact excitation effective collision strengths $\Upsilon$
   of $\mathrm{Ni}^{3+}$ for a Maxwellian electron distribution.}
\label{fig:ni3ups}
\end{figure*}

The collision strength $(\Omega_{ij})$ between an initial state $i$ and a final 
state $j$ is directly related to the electron-impact collisional excitation cross 
section $\sigma_{ij}$ by 
\begin{equation}
  \sigma_{ij}\ =\ \Omega_{ij}\,\frac{\pi a_{0}^2}{\omega_{i}k_{i}^2}\,,
\label{eq:colstr}
\end{equation}
where $\omega_{i}=2J_{i}+1$ is the statistical weight of level $i$ 
and $k_{i}^2$ is the incident electron energy in Rydberg. 
In the majority of astrophysical plasmas the electron velocity distribution is 
Maxwellian for a certain temperature $T$. 
Hence to aid plasma modeling we have performed a convolution of the collision 
strength $\Omega_{ij}$ in terms of the Maxwellian distribution to obtain the 
associated effective collision strengths $\Upsilon_{ij}$.
\begin{equation}
  \Upsilon_{ij}\ =\ \int_{0}^{\infty} \Omega_{ij}(E_{j})\, 
  \exp \left( - \frac{E_{j}}{kT} \right)\,
  \rd \left(\frac{E_{j}}{kT} \right)
  \label{eq:maxwell}
\end{equation}
where $T$ is the Maxwellian electron temperature in $\kelvin$, 
$E_{j}$ is the final energy of the incident electron and $k$ is Boltzmann's constant.
For high temperatures the Maxwellian has a long tail and it is necessary to calculate 
$\Omega_{ij}$ up to a suitably high energy. 
This requires the inclusion of a large number of continuum states in the 
Hamiltonian and thus increase the corresponding size of the matrices beyond the 
computation capabilities available. 
To alleviate this problem we have calculated, using {\sc darc}, the infinite-energy limit 
for the electric dipole transitions and interpolated $\Omega_{ij}$ in the scaled 
Burgess-Tully domain~\cite{burgess1992} between the last energy calculated in the 
outer region with {\sc pstgf} and the infinite energy point. 
For the present calculation we can therefore guarantee accuracy of 
the $\Upsilon_{ij}$ up to the order of $5.5 \Ry$, equivalent 
to $2 \times 10^{6} \kelvin$. 
The temperature of maximum-abundance for $\mathrm{Ni^{3+}}$ is approximately
$5 \times 10^4 \kelvin$ \cite{mazzotta1998,bryans2006}, 
which indicates that the present evaluation is sufficient to model and resolve 
the emission features of the \ion{Ni}{iv} lines in the range of temperatures 
where $\mathrm{Ni^{3+}}$ is abundant.
We have used the program 
{\sc adasexj} (Griffin and Badnell, unpublished)
to perform the convolution of the $\Omega$ and calculated the
Maxwellian $\Upsilon$.
We create a level-resolved specific ion {\sc adf04} file to store all the relevant
collision-radiative parameters.
This {\sc adf04} file can be used as standard input to usual collision-radiative
modelling software, for example the ADAS series 2 \citep{summers1994}.
  
In the present work we have computed collision strengths $\Omega_{ij}$ and 
effective collision strengths $\Upsilon_{ij}$ for the electron-impact excitation 
of the $\mathrm{Ni}^{3+}$ ion for transitions between the lowest 262 levels, 
a total of $34\,191$ forbidden and allowed lines.
The highest energy considered was $5.5 \Ry$, adequate when compared to the 
ionization energy of $4.037 \Ry$ \cite{nist2018}.
Above this ionization energy, the collision strengths follow an asymptotic
behaviour and can be interpolated with the infinite energy limit point in
the Burgess-Tully domain~\cite{burgess1992}. 

In Figure \ref{fig:ni3omg}, we present the collision strength $\Omega_{ij}$ for 
the electron-impact excitation of some selected transitions of 
the $\mathrm{Ni}^{3+}$ ion. 
For all transitions we observe the expected series of resonances in the low 
energy region converging onto the target state thresholds included in the 
CC expansion, and a background above this that depends on the type of transition 
considered.
The most useful transitions for astrophysical diagnosis are the M1 transitions 
between the levels of the ground term. 
A peculiarity of the present system is that the first levels with odd parity are 
highly excited, the first one listed as level 65. 
As a consequence of this the electric dipole E1 allowed transitions from the 
ground term are paradoxically very weak in comparison with the other M1 and E2 
transitions within the lower-excited levels, 
in fact it is in this transition where both versions of the calculation, with
shifted and unshifted target energies, disagree the most (pannel (g)).
The cause of this disagreement is that in both versions of the calculation, 
the wave functions of the atomic states have not been modified,
but the energies have, since in one of them they have been shifted to the 
recommended values of NIST.
As a consequence, the line strengths $S$ have the same value in both calculations.
At high energies, the collision strengths are determined by the infinite 
energy point, which in the case of E1 transitions depends only on the value of $S$, 
see \citet{burgess1992}.
In Figure \ref{fig:ni3omg} it is appreciated that for the E1 transition $1-65$ 
the collision strengths obtained using both versions disagree at low impact
energies, but they converge at high ones.

We present in Figure \ref{fig:ni3ups}  the corresponding Maxwellian averaged 
effective collision strengths $\Upsilon_{ij}$ for the same transitions depicted 
in Figure \ref{fig:ni3omg}, for a range of electron 
temperatures $T_{e} = 10^{3} - 10^{6} \kelvin$. 
Clearly evident is the strong enhancement of the collision rates due to the 
proper delineation of the Rydberg resonance features in the collision strengths. 
For the online material we provide tables of the calculated effective collision
strengths for all the $34\,191$ transitions between all levels 
of $\mathrm{Ni^{3+}}$.
For non-Maxwellian modeling or for any application that requires the direct
collision strengths we direct the reader to our public 
ftp server \footnote{\url{http://web.am.qub.ac.uk/wp/apa/publications-data/}}.
We also refer to the 
OPEN-ADAS \footnote{\url{http://open.adas.ac.uk}}
data base for the general {\sc adf04} file.

As a convergence test we have compared different {\sc adf04} files,
in the first one we have included in the partial wave expansion angular momenta
up to $J=30$ and no top-up,
in a second one we have added the top-up to the $J=30$ expansion, 
and finally our recommended data with the partial wave expansion extended
up to $J=36$ plus top-up.
The largest differences remain between the versions with and without top-up,
in that case the average difference between all the transitions values $0.5\%$.
In particular for the E1 allowed transitions, the maximum difference reaches 
the $100\%$, while for the forbidden transitons this maximum difference is 
of the order $10\%$.
If we add the top-up to the $J=30$ expansion the differences reduce significantly,
the average difference is reduced to the $0.02\%$,
and the maximum difference for the E1 transitions to the $33\%$, 
and only in six E1 transitions is above the $10\%$,
these six transitions are between very excited states, above 100, 
and they are irrelevant for the modeling.
It is clear the calculation is properly converged in terms of the expansion
in partial waves once the top-up is added,
expansion up to $J=30$ and $J=36$ produce equal results.

\subsection{Photoionization of $\mathrm{Ni^{2+}}$}
\label{subsec:Ni2pi}

\begin{figure*}
   \includegraphics[width=0.4\textwidth]{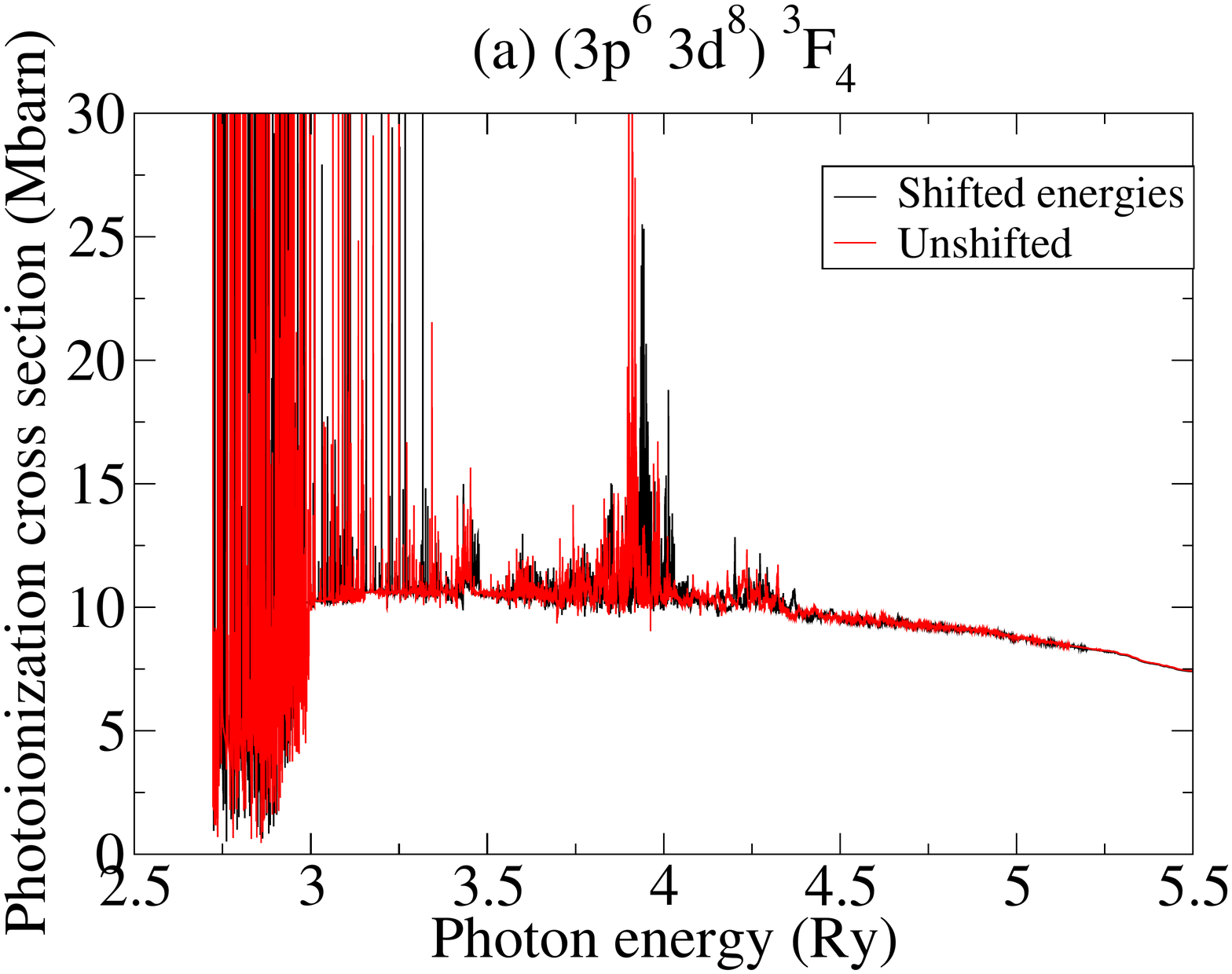} \,
   \includegraphics[width=0.4\textwidth]{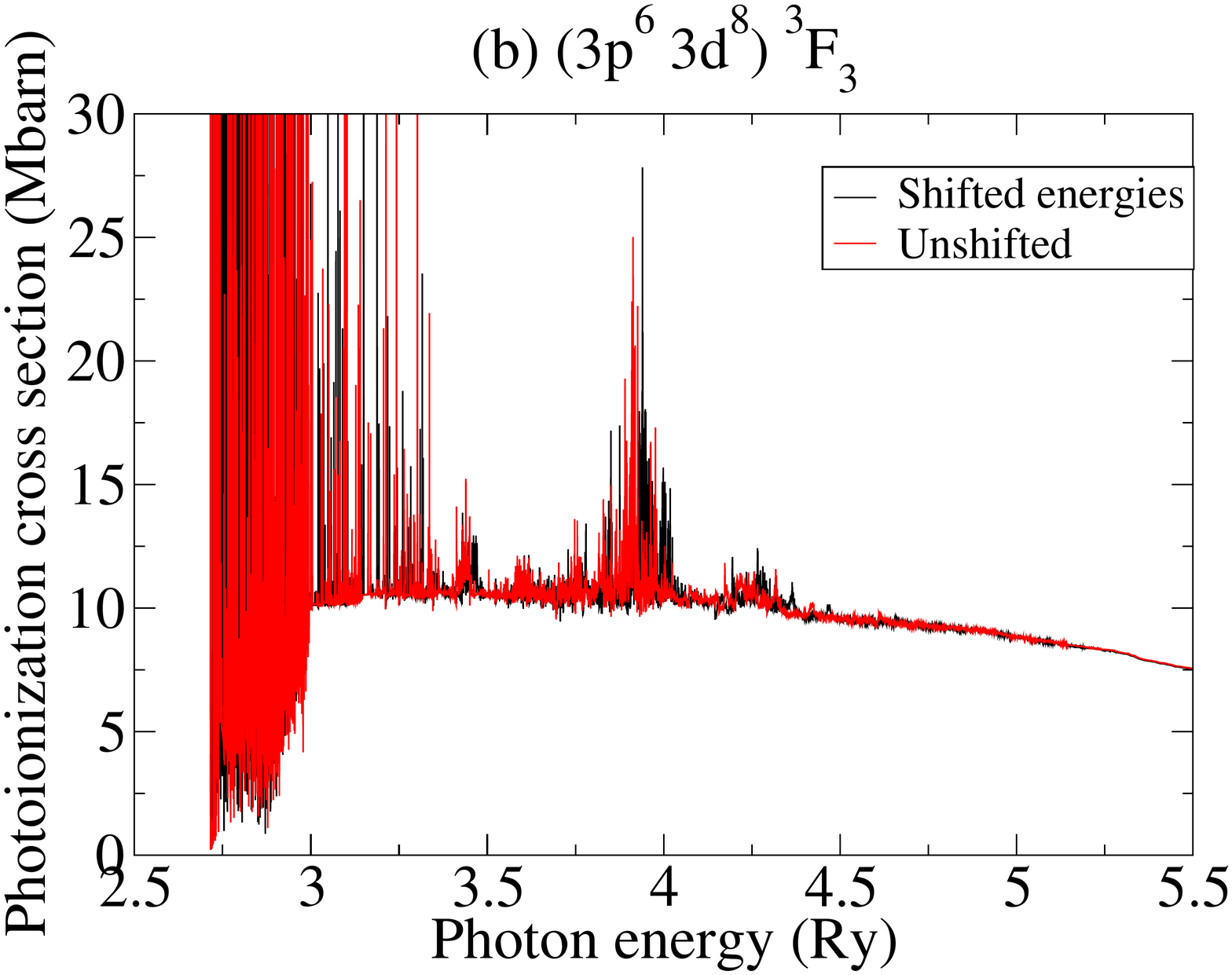} \\
   \includegraphics[width=0.4\textwidth]{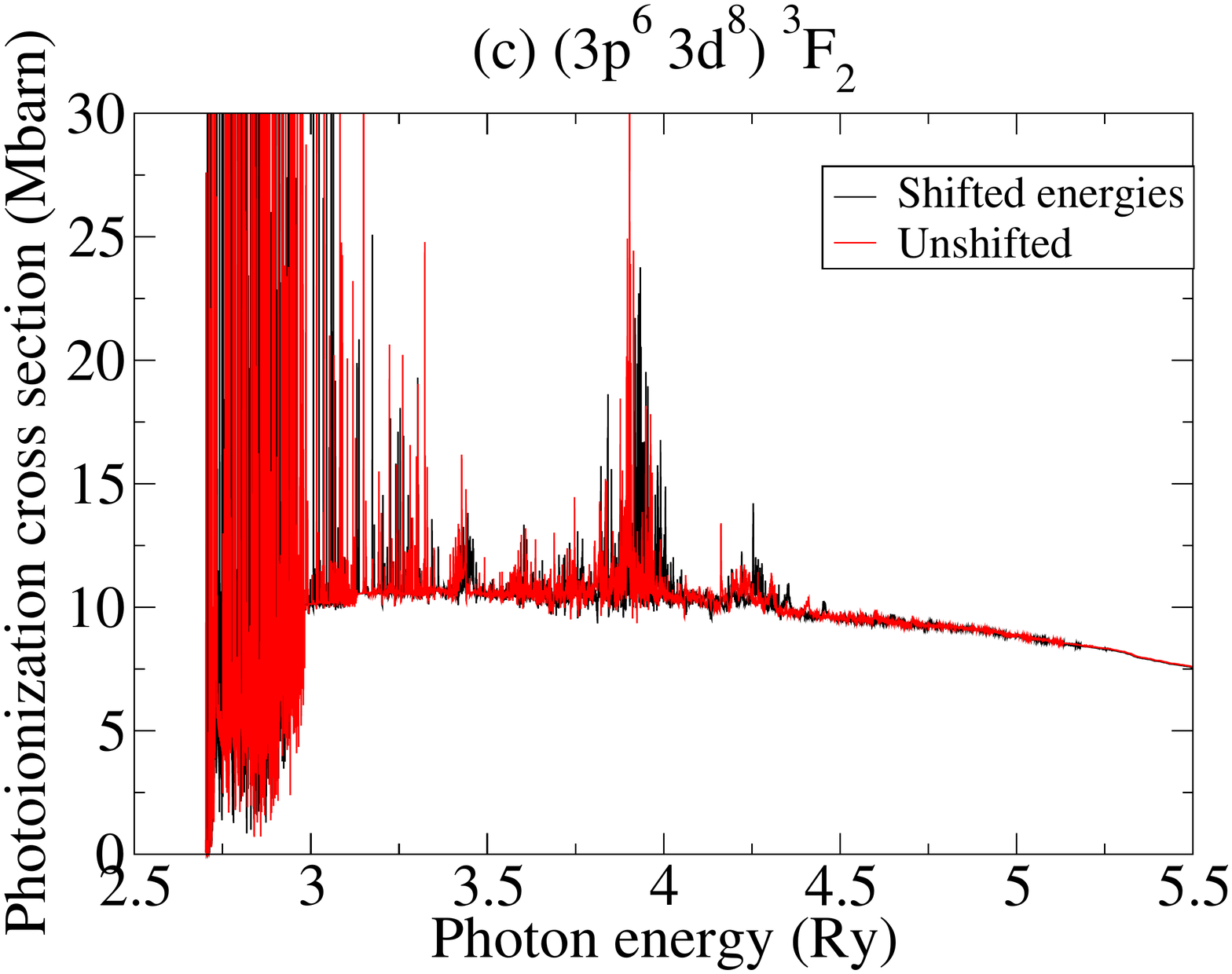} \,
   \includegraphics[width=0.4\textwidth]{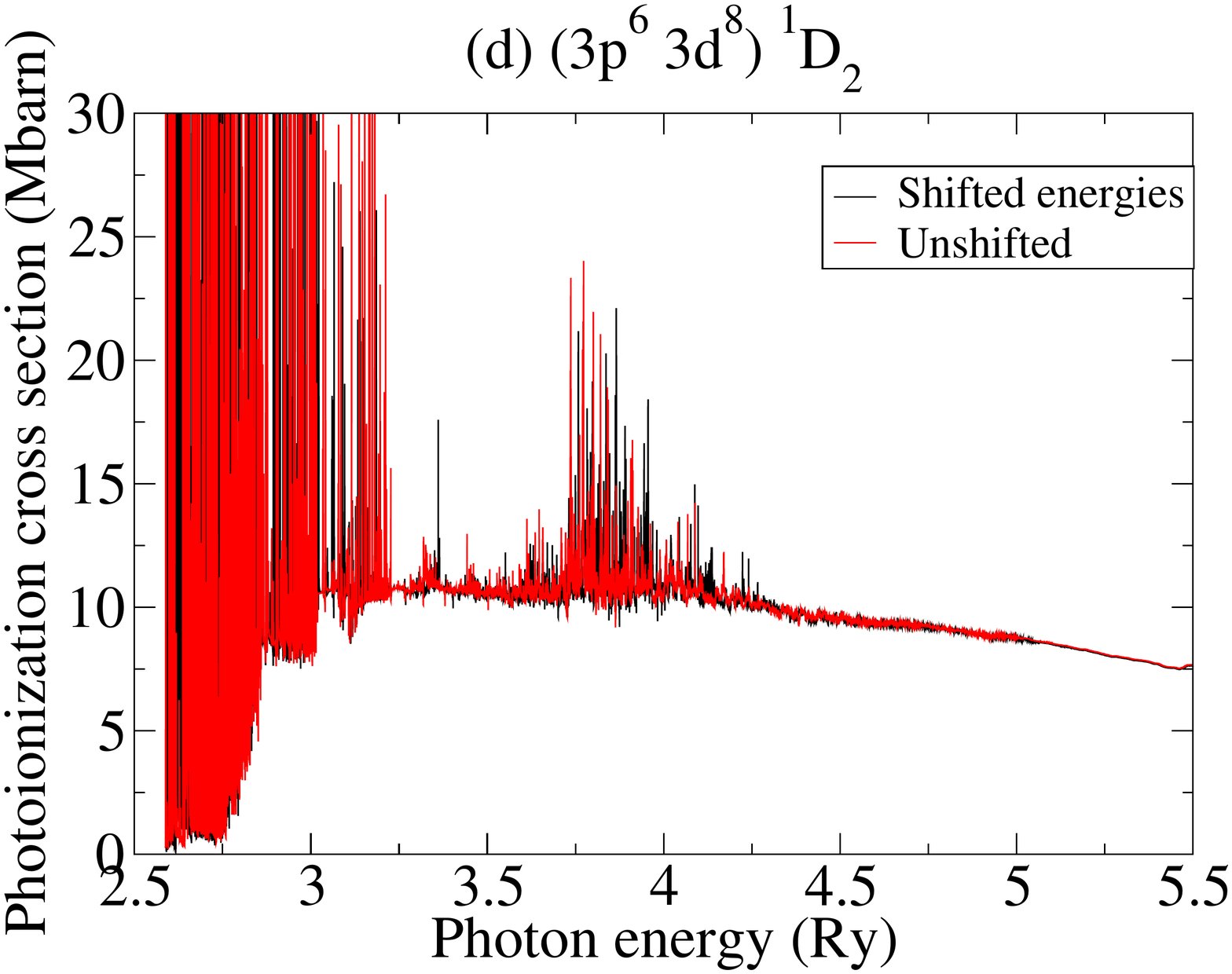} \\
   \includegraphics[width=0.4\textwidth]{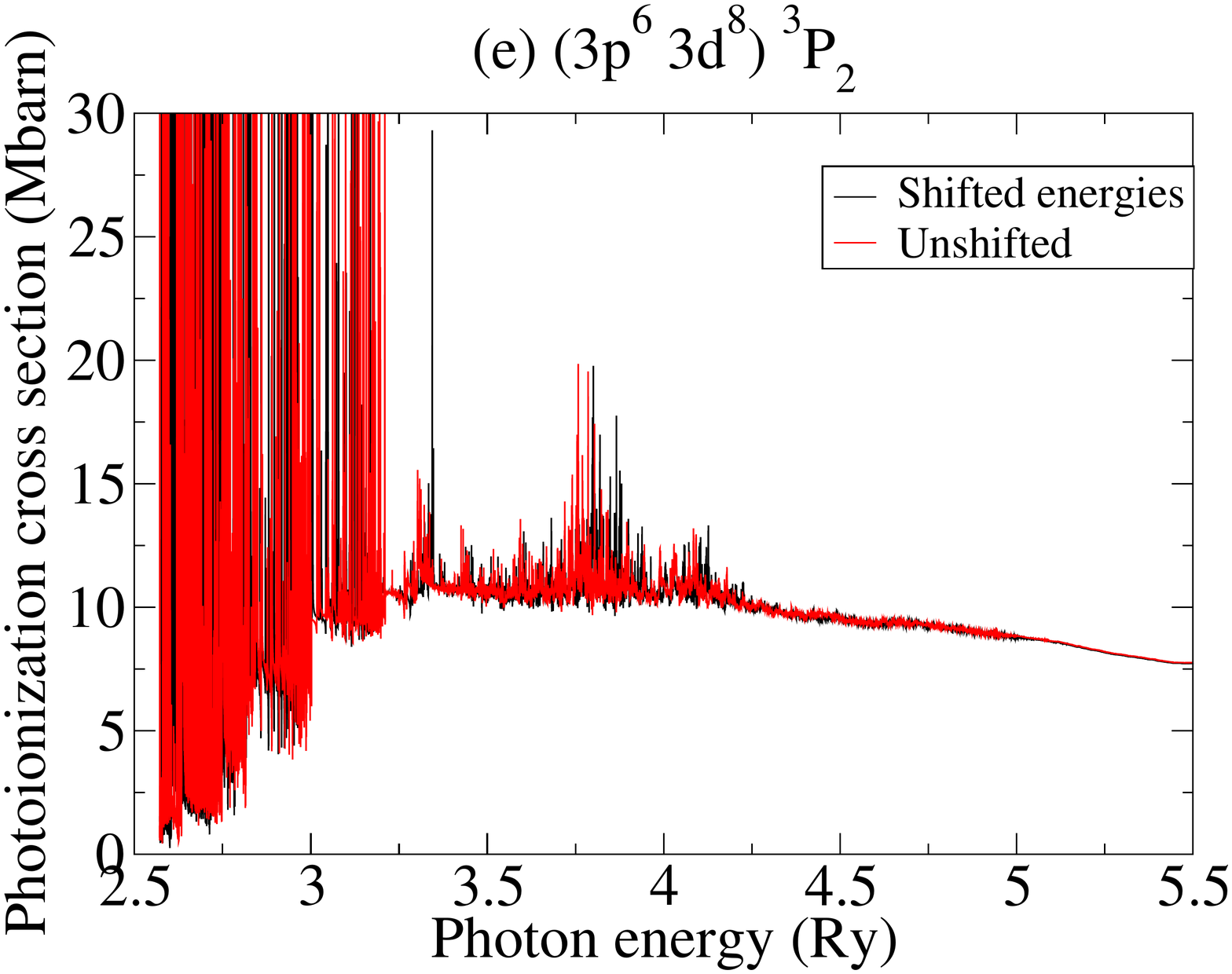} \,
   \includegraphics[width=0.4\textwidth]{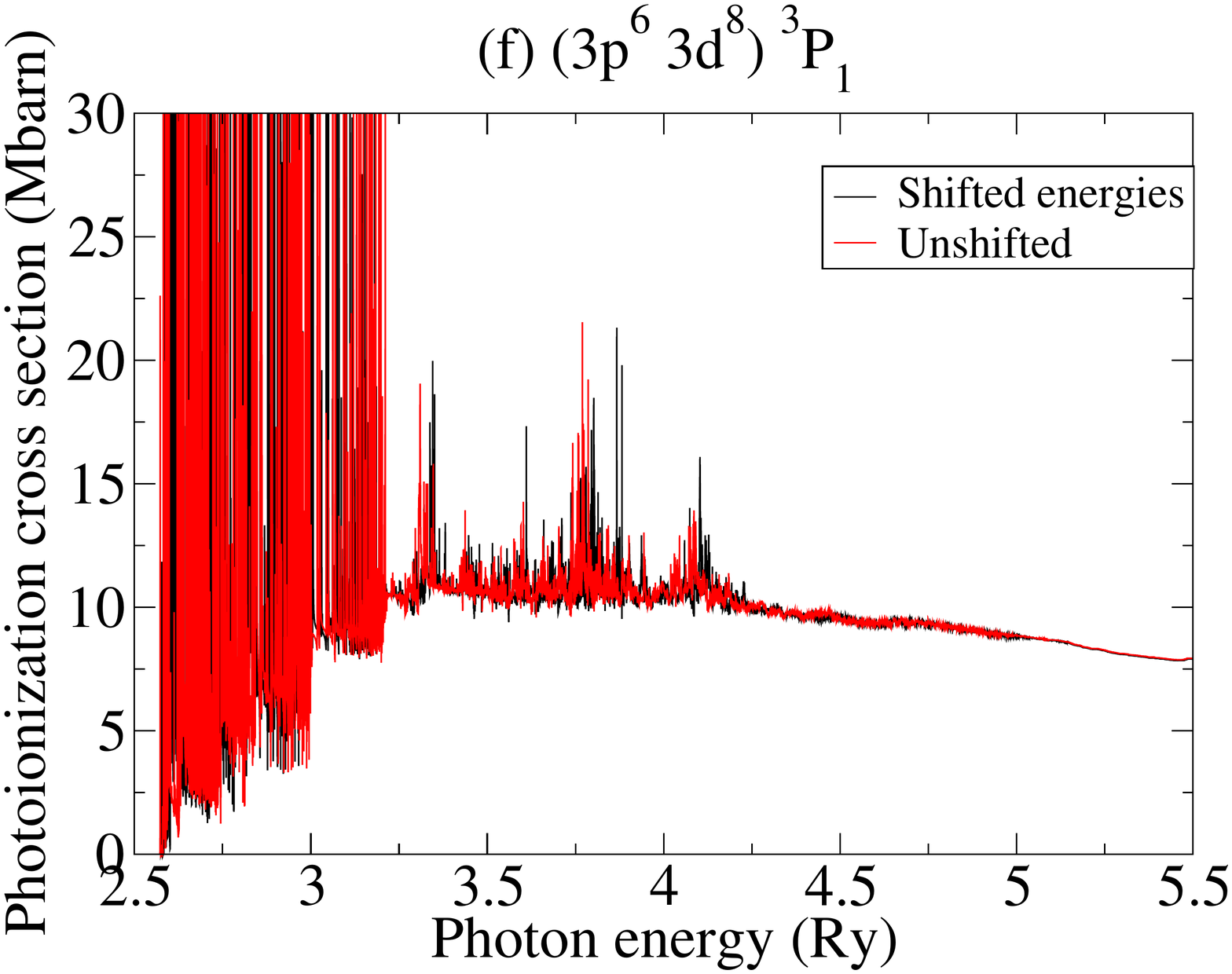} \\
   \includegraphics[width=0.4\textwidth]{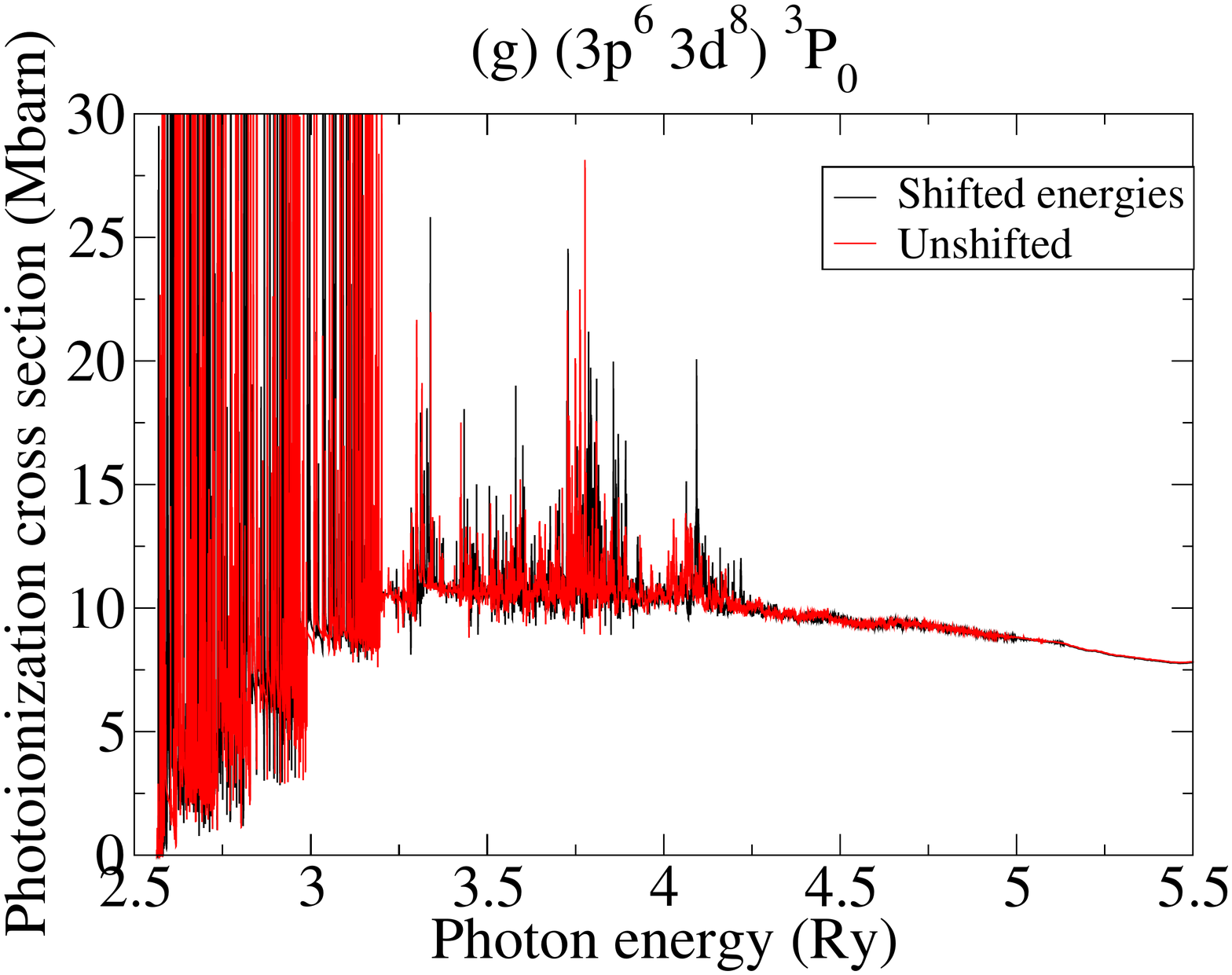} \,
   \includegraphics[width=0.4\textwidth]{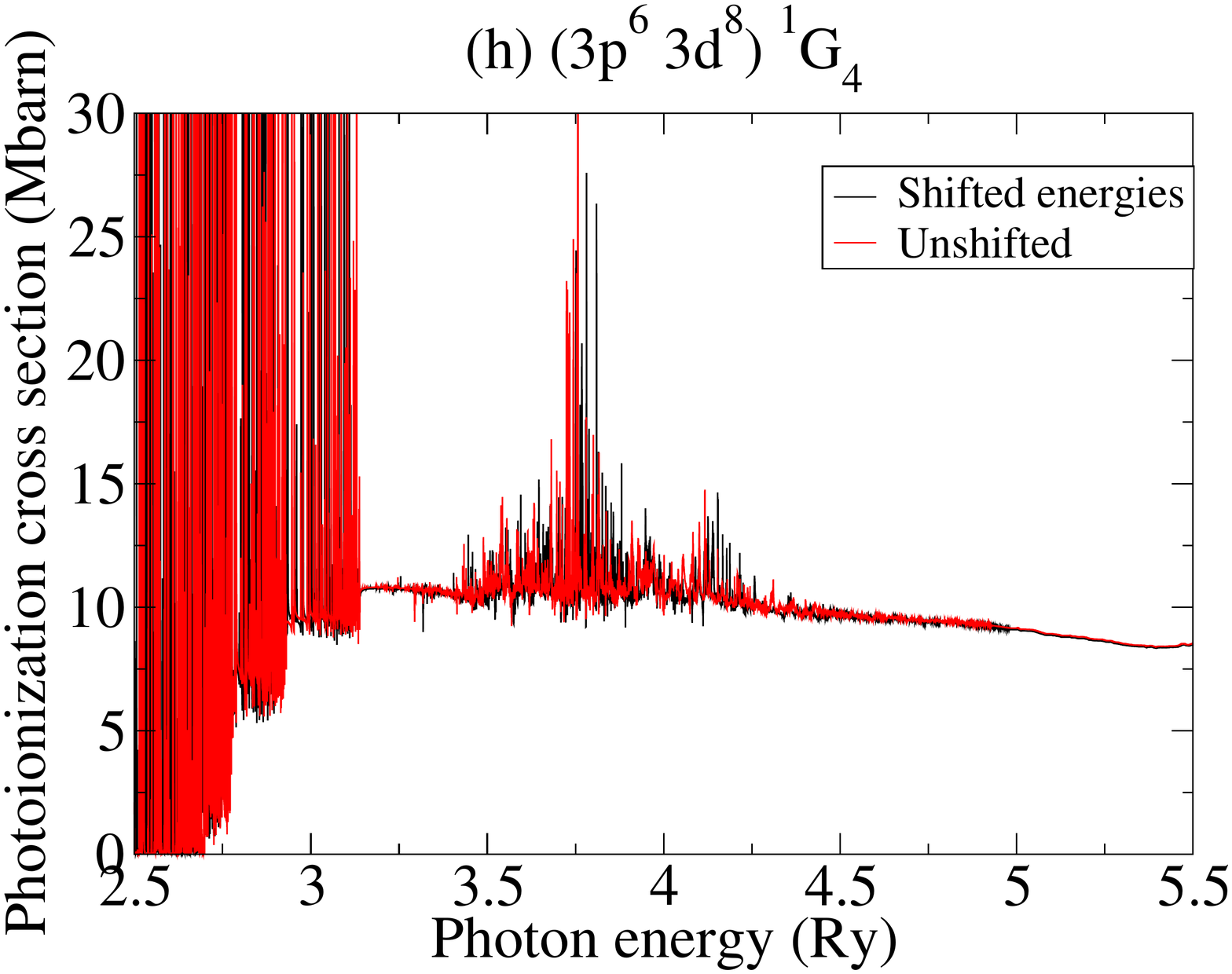} 
\caption{Photoionization cross sections versus the photon energy for
   $\mathrm{Ni}^{2+}$ from ground lowest-excited initial states.
   Colour online.}
\label{fig:ni2pi}
\end{figure*}

We have calculated level resolved photoinization cross sections 
of $\mathrm{Ni}^{2+}$ from its 20 lowest-energy levels 
for each $J^{\pi}$ symmetry with $J=0-4$ and even parity,
to all the 262 final states of $\mathrm{Ni}^{3+}$ included in the CC expansion.
In a stellar cloud most of the population of $\mathrm{Ni}^{2+}$
will occupy the ground level $\mathrm{3p^6\,3d^8\,^3F_4}$ with only
a small fraction populating the metastable levels of the ground term
$\mathrm{^3F_3}$, $\mathrm{^3F_2}$ and $\mathrm{^1D_2}$.
In an usual stellar cloud these metastable levels contribute to the opacity 
much less than the ground state. 
In addition, the $\mathrm{^1D_2}$ level is coupled to the $\mathrm{^3F}$ term 
through a spin-changing M1 / E2 transition with a very small transition probability.
Table \ref{tab:Ni2_A} shows the Einstein transition coefficients 
for transitions among the three first terms of $\mathrm{Ni}^{2+}$
taken from \citet{garstang1958}. 
Clearly transitions between these levels are very weak, with A-values of the 
order of $10^{-1}-10^{-2} \mathrm{s}^{-1}$. 
These levels can therefore be considered as metastable when included in an 
opacity model. 
The first odd level of $\mathrm{Ni}^{2+}$ is the $\mathrm{3p^6\,3d^7\,4p\,^5F_5}$ 
state, with an excitation energy of $110\,213\,\mathrm{cm}^{-1}$ with respect 
the ground state, see \cite{nist2018}.
All the levels below it are radiatively connected to the ground and metastable
states through one or several forbidden E2 and M1 transitions.
These transitions are known to be more intense as the level-energy difference 
is greater and hence terms above $\mathrm{^1G}$ will not be populated in a 
low-density cloud.

\begin{table}
\caption{Spontaneous emission coefficients for transitions between the 
   lowest-excited levels of $\mathrm{Ni}^{2+}$ $\mathrm{3p^6\,3d^8}$}
\label{tab:Ni2_A}
\centering
\begin{tabular}{llcrr}
   \hline
   Lower            & Upper            &         &            &     \\
   level            & level            & Type    & WL         & $A$ \\
   \hline
   $\mathrm{^3F_4}$ & $\mathrm{^1D_2}$ & E2      & $7\,124.8$ & $4.5\,[-3]$ \\
   $\mathrm{^3F_3}$ & $\mathrm{^1D_2}$ & M1 / E2 & $7\,889.9$ & $4.8\,[-1]$ \\
   $\mathrm{^3F_2}$ & $\mathrm{^1D_2}$ & M1 / E2 & $8\,499.6$ & $2.1\,[-1]$ \\
   $\mathrm{^3F_4}$ & $\mathrm{^3P_2}$ & E2      & $6\,000.2$ & $5.0\,[-2]$ \\
   $\mathrm{^3F_3}$ & $\mathrm{^3P_1}$ & E2      & $6\,401.5$ & $3.8\,[-2]$ \\
   $\mathrm{^3F_3}$ & $\mathrm{^3P_2}$ & M1 / E2 & $6\,533.8$ & $1.1\,[-1]$ \\
   $\mathrm{^3F_2}$ & $\mathrm{^3P_0}$ & E2      & $6\,682.2$ & $4.6\,[-2]$ \\
   $\mathrm{^3F_2}$ & $\mathrm{^3P_1}$ & M1 / E2 & $6\,797.1$ & $1.6\,[-2]$ \\
   $\mathrm{^3F_2}$ & $\mathrm{^3P_2}$ & M1 / E2 & $6\,946.4$ & $2.3\,[-2]$ \\
   $\mathrm{^1D_2}$ & $\mathrm{^3P_0}$ & E2      & $31\,259$  & $2.4\,[-6]$ \\
   $\mathrm{^1D_2}$ & $\mathrm{^3P_1}$ & M1 / E2 & $33\,942$  & $9.0\,[-2]$ \\
   $\mathrm{^1D_2}$ & $\mathrm{^3P_2}$ & M1 / E2 & $38\,023$  & $9.8\,[-2]$ \\
   \hline
\end{tabular}
\\
   \flushleft{Key: WL, wavelength in air ($\AA$); 
   $A$, Einstein spontaneous emission coefficient $\mathrm{s}^{-1}$,
   $A\,[B]$ denotes $A \times 10^B$. \\
   Data from \citep{garstang1958}.}
\end{table}

In Figure \ref{fig:ni2pi} we present the total photoionization cross section 
of $\mathrm{Ni}^{2+}$ from its ground state as a function of photon energy in $\Ry$,
as well as the seven lowest metastable levels.
The cross section depicts a typical structure of large Rydberg resonances on a 
continuous background.
In order to reproduce the high-energy region above approximately $5.5 \Ry$ it is 
necessary to include more continuum functions and additional highly-excited levels  
in the CC expansion of the target.
In the online material we provide a full table of fully resolved photoinization 
cross sections from the 20 lowest-excited levels of with $J=0-4$ and even parity
of $\mathrm{Ni}^{2+}$ to the 262 lowest-excited levels of $\mathrm{Ni}^{3+}$.
These cross sections can be considered of high-quality for photon energies
up to $5.5 \Ry$ and can be used for any opacity model.

\begin{figure}
  \includegraphics[width=\columnwidth]{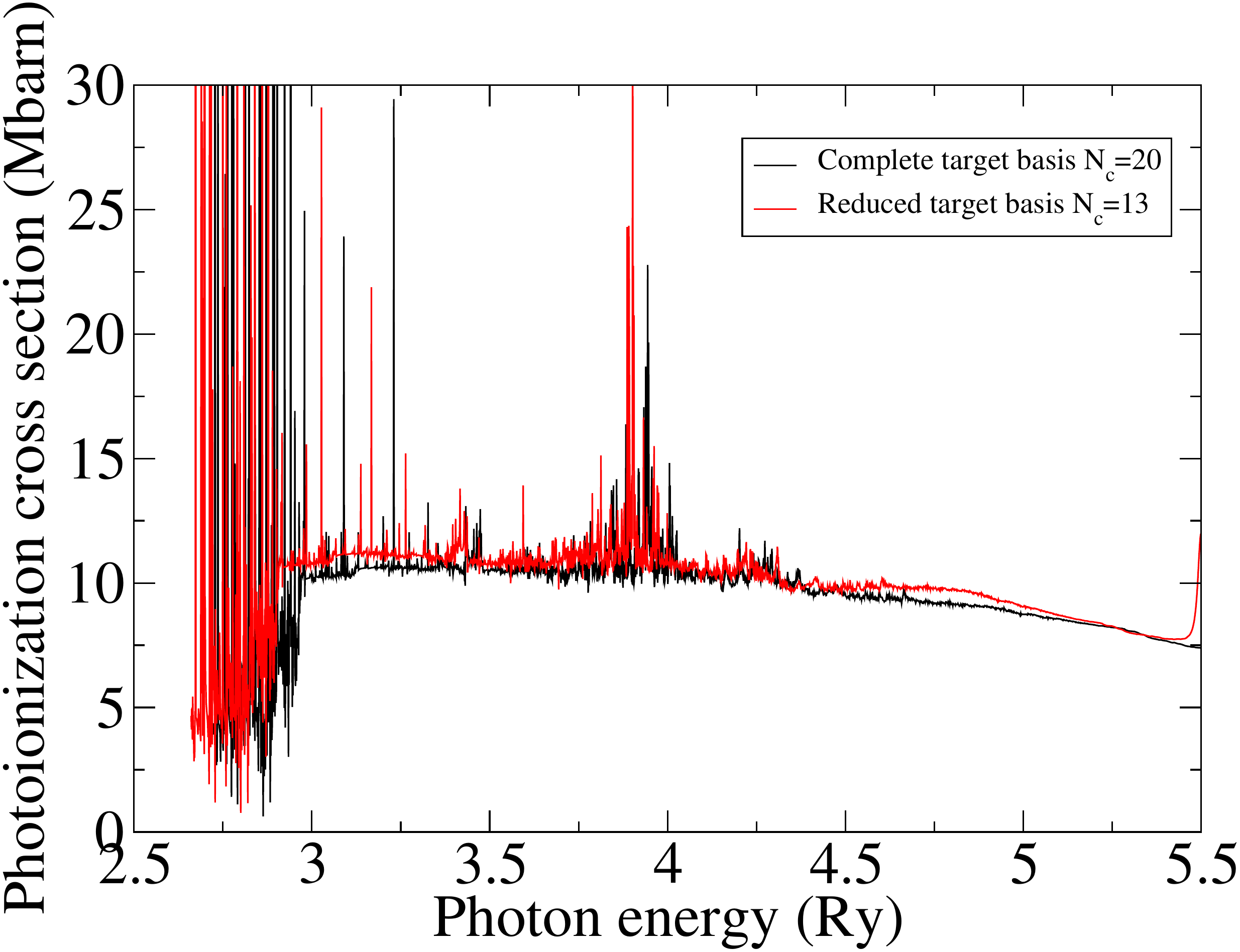}
\caption{Photoionization cross section of $\mathrm{Ni^{2+}}$ ion initially
   in its ground state $\mathrm{3p^6\,3d^8\,^3F_4}$ 
   with two different expansions.
   Black line: calculation with the whole set of configurations and $N_c=20$;
   Red line: calculation with a truncated set of configurations and $N_c=13$.}
\label{fig:ni2pi_comp0}
\end{figure}

In Figure \ref{fig:ni2pi_comp0} a test of convergence for the calculation is presented.
We compare two calculations performed with the same atomic structure of the target.
In the first one (black line) we included in the configuration basis set of the
($N+1$)-electron system all the configurations derived from the addition of one extra
electron into all the available orbitals included in the expansion to those 
configurations listed in Table \ref{tab:confs} and with an expansion of the continuum 
including $N_c=20$ functions.
In the second one (red line) the configuration set was reduced somewhat extracting 
from Table \ref{tab:confs} the $\mathrm{3p^5\,3d^7\,5s}$, $\mathrm{3p^5\,3d^7\,5p}$, 
$\mathrm{3p^5\,3d^7\,6s}$, $\mathrm{3p^5\,3d^7\,6p}$, 
$\mathrm{3p^6\,3d^5\,4d^2}$, $\mathrm{3p^6\,3d^6\,4d}$, $\mathrm{3p^4\,3d^9}$ 
basis configurations to build the ($N+1$)-electron system expansion,
and with $N_c=13$ functions for the expansion of the continuum.
Evidently, there is a very small difference between both calculations with respect 
to the background, the position of the resonances and their heights.
We can be confident therefore that the present calculation has converged with 
regard to the target description, the size of the continuum basis and the mesh 
size adopted in the low-energy region.
For higher photon energies above $5.5 \Ry$ a similar guarantee of the accuracy 
of the cross sections cannot be made due to the effect of additional excited levels 
which are not included in our CC expansion. 

\begin{figure}
   \includegraphics[width=\columnwidth]{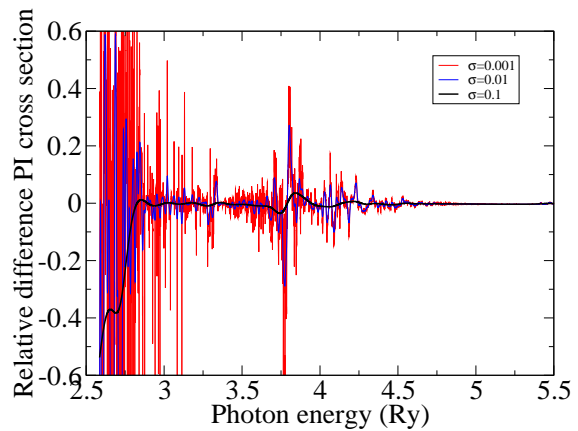}
\caption{Relative difference of the photoionization cross section 
   of $\mathrm{Ni}^{2+}$ from its ground level
   $\mathrm{3p^6\,3d^8\,^3F_4}$
   between shifted and unshifted versions of the calculation
   after several Gaussian convolutions with different widths.
   Colour online.}
\label{fig:3F4_cnvl}
\end{figure}

As a test of accuracy we investigate in Figure \ref{fig:3F4_cnvl} the relative 
difference of the photoionization cross sections produced when the target levels 
are shifted to their exact observed positions or left unshifted as 
the {\it ab initio} values.
Relative differences of 
\begin{equation}
   \delta=\frac{\sigma_{sh}-\sigma_{un}}{\sigma_{sh}}
\label{eq:delta}
\end{equation}
where $\sigma$ represents the convoluted cross section with a Gaussian enveloping
for several widths. 
The largest deviation occurs for the lower photon energies.
For those energies the difference in the positioning of the resonances is 
the dominant contribution to the global error.
For photon energies above $3 \Ry$, 
equivalent to wavelengths shorter than $303.76\,\AA$,
the relative deviation for the convolution with width $10^{-2}\Ry$ 
remains below the $10\%$ level in almost the entire domain. 
At a photon energy of $E=3.81 \Ry$ the deviation reaches a maximum of $28\%$,
just for a single resonance.
For a convolution width of $10^{-3} \Ry$ and photon energies above $3 \Ry$
the relative difference remains below the $20\%$ threshold.
We estimate the accuracy of the present data to be approximately $20\%$ in the 
worst case for wavelengths in the ultraviolet, above the ionization 
limit of~$\mathrm{Ni}^{2+}$.

\section{Modeling of diagnostics.}
\label{sec:modeling}

\begin{figure*}
   \includegraphics[width=0.48\textwidth]{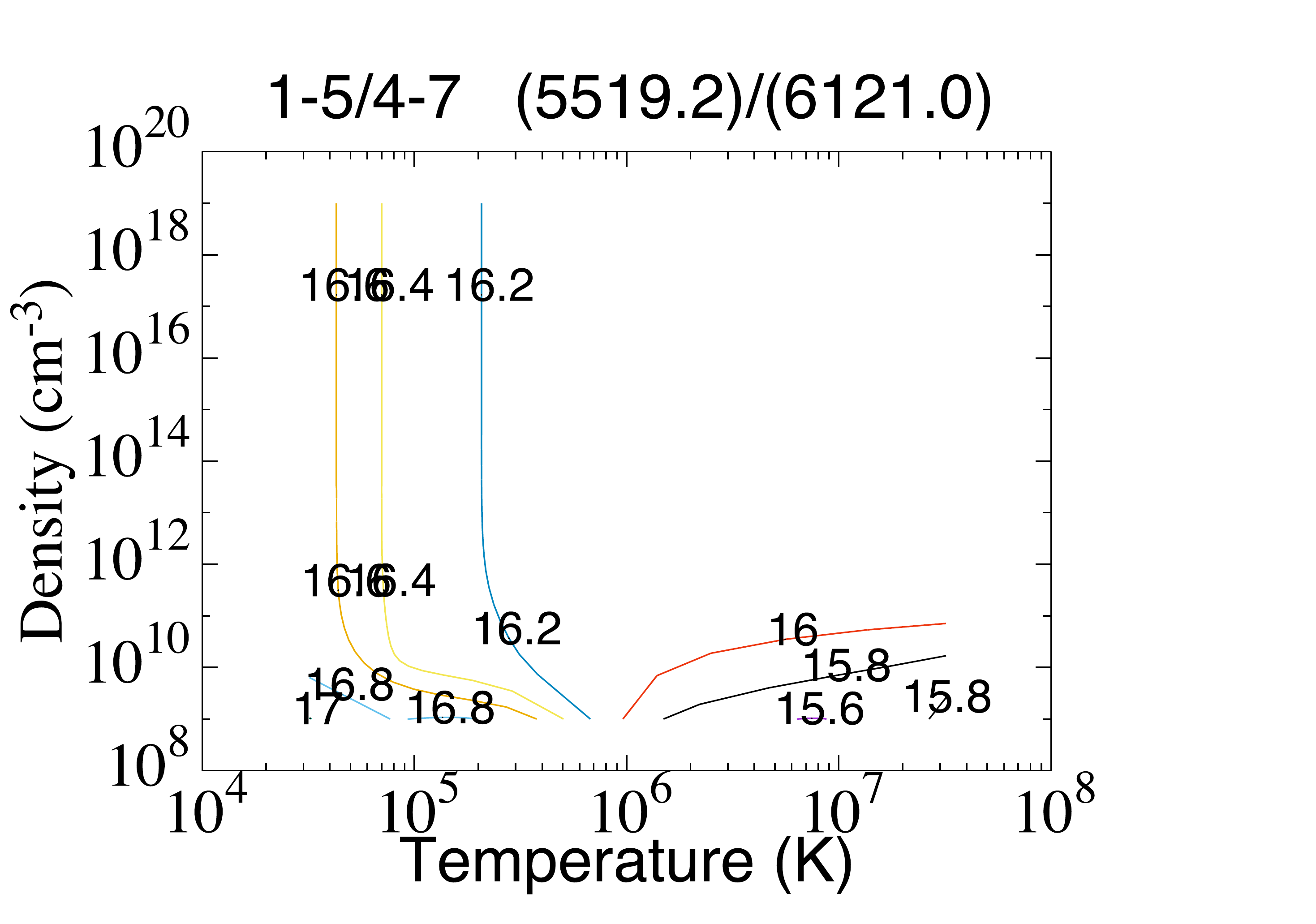} \,
   \includegraphics[width=0.48\textwidth]{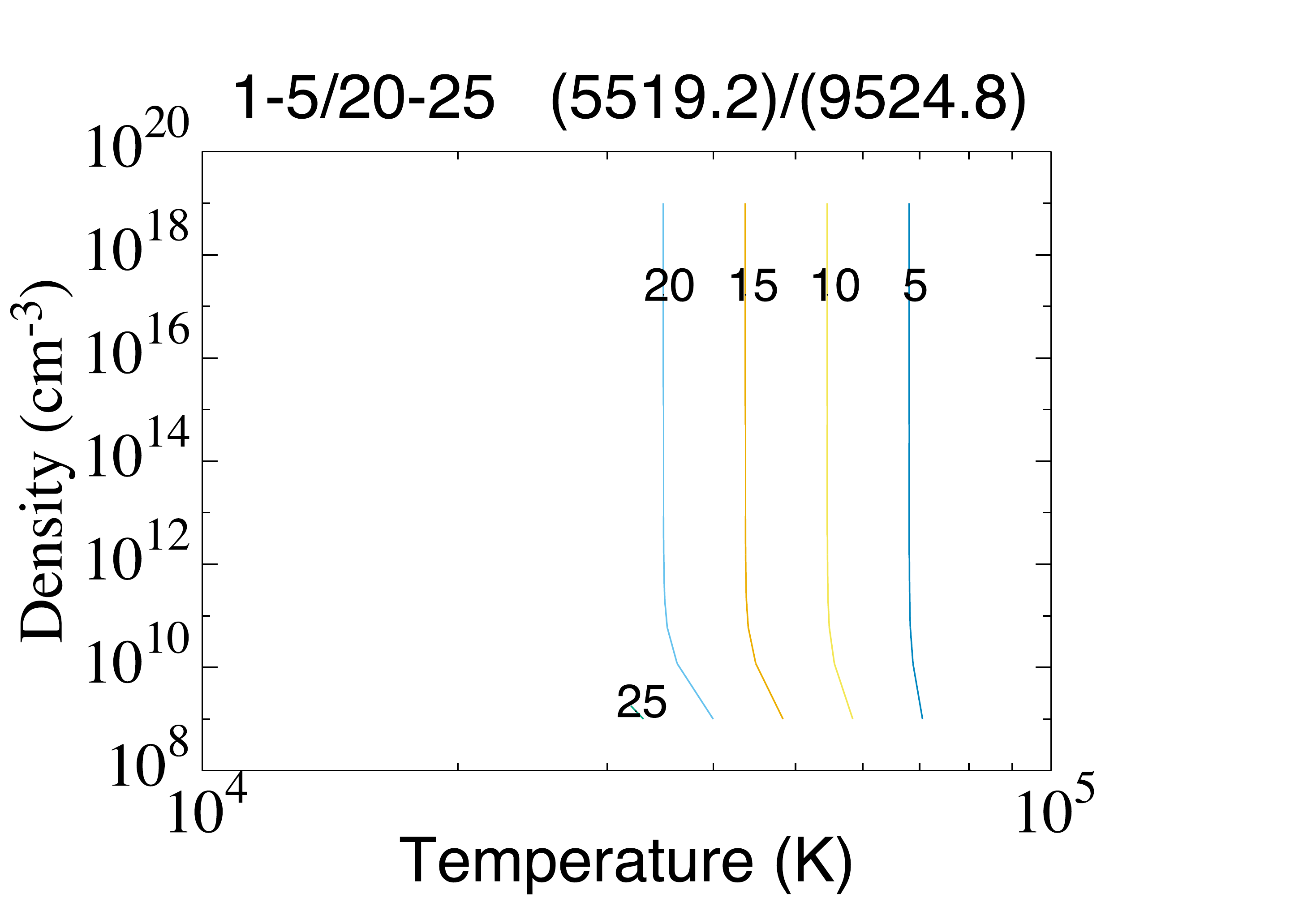} \\
   \includegraphics[width=0.48\textwidth]{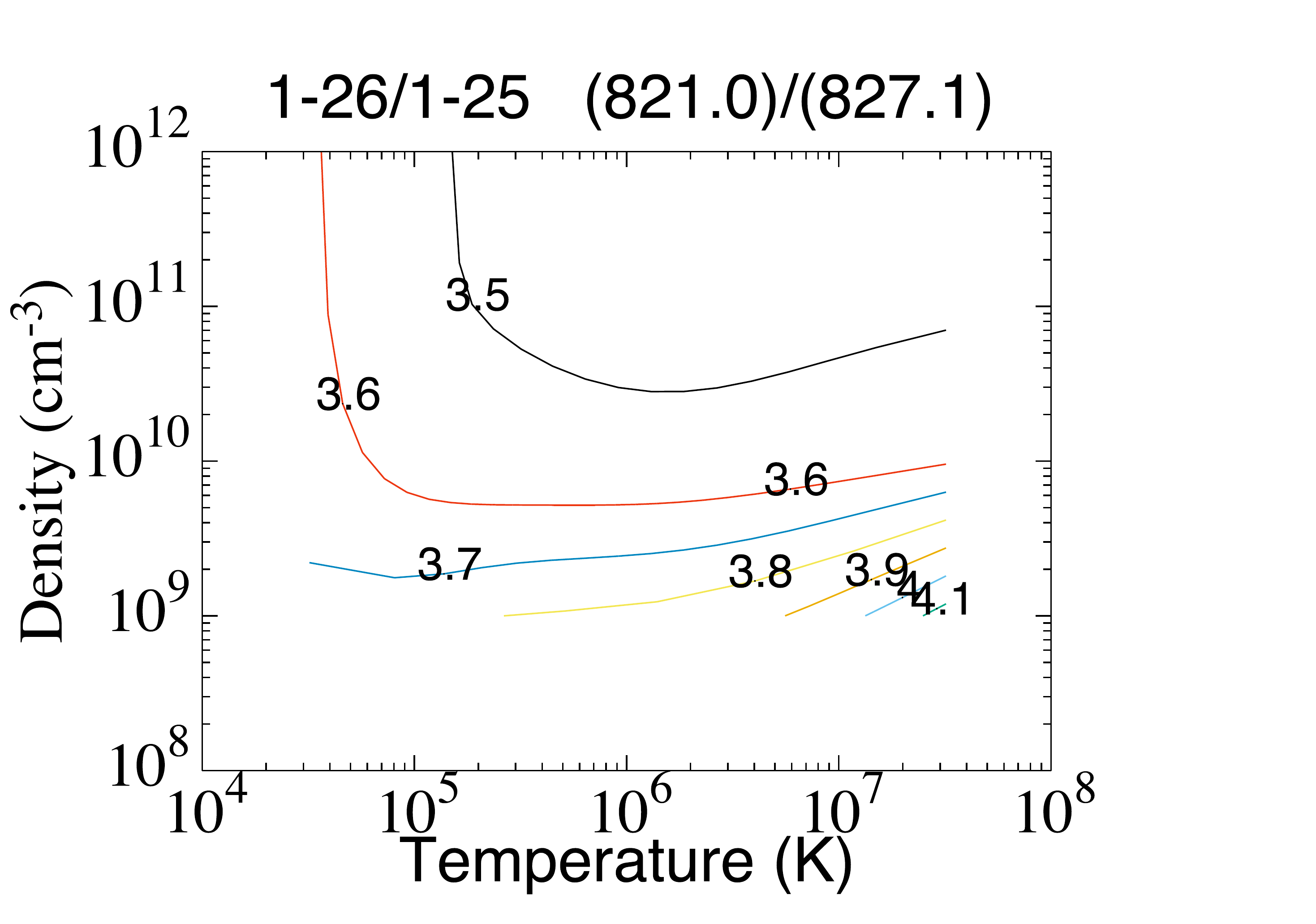} \,
   \includegraphics[width=0.48\textwidth]{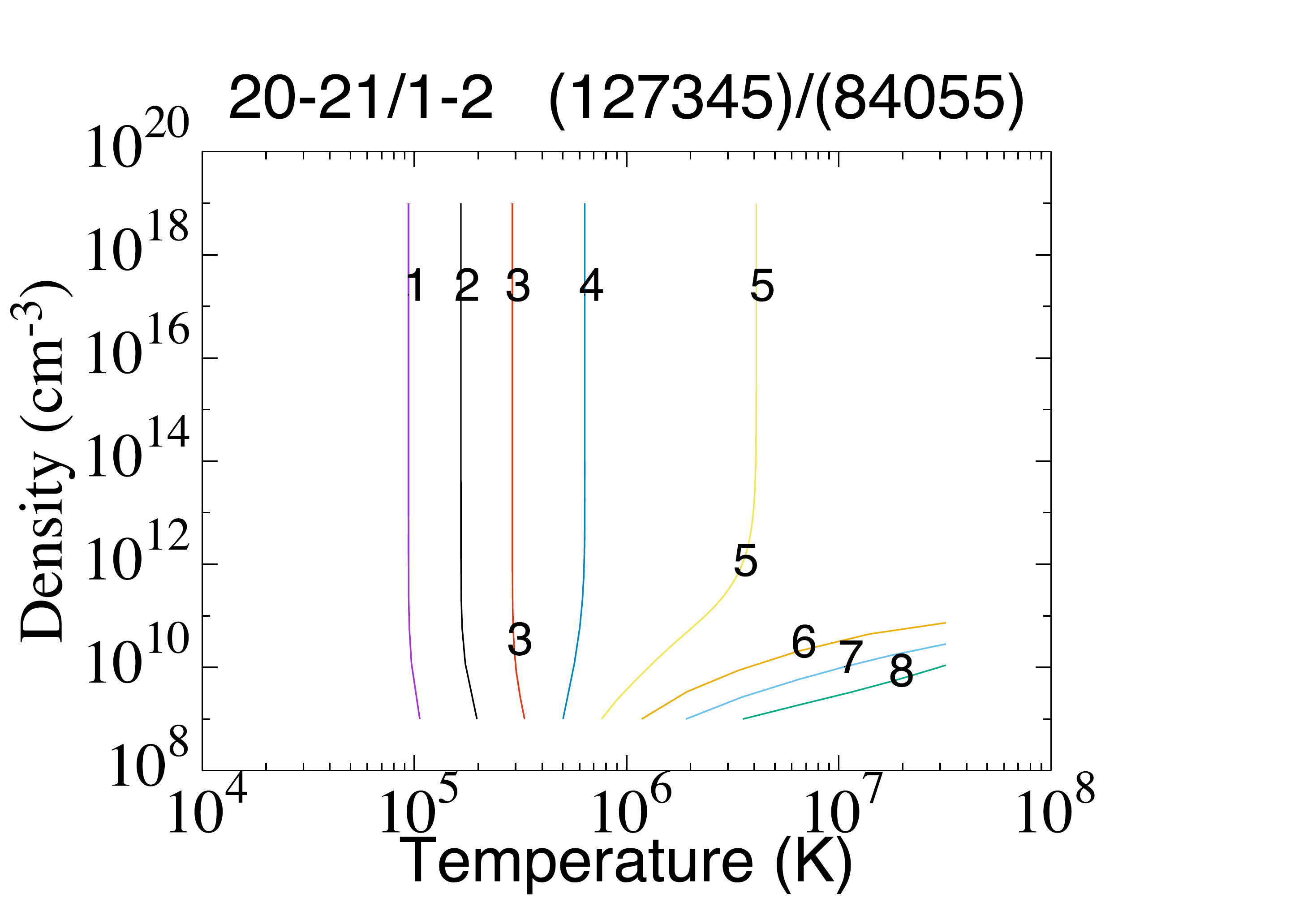} 
\caption{Line intensity ratio $\frac{I_{\lambda 1}}{I_{\lambda 2}}$ 
   versus electron temperature and density
   for some selected pairs of lines of \ion{Ni}{iv}.
   Colour online. }
\label{fig:linrat}
\end{figure*}

With the calculated effective collision strengths for the 
electron-impact excitation of $\mathrm{Ni}^{3+}$
we have performed a collision-radiative model.
We use the program {\sc colrad},
which calculates the line intensities from the radiative transition probabilities 
and effective collision strengths stored in the {\sc adf04} file.
For low densities, the only mechanism of population is the collisional
excitation from the ground or a metastable state, 
following radiative-decay cascade.
In Table \ref{tab:linrat} we have selected four line ratios to check their 
validity as diagnostics.
These transitions were considered in a previous calculation by \cite{pradhan1993b}
for the isoelectronic ion $\mathrm{Fe}^{+}$, and hence provide a benchmark for 
the current analysis.

\begin{table}
\caption{\ion{Ni}{iv} line ratios used for plasma diagnostics}
\label{tab:linrat}
\begin{tabular}{ccrccr}
   \hline
   \multicolumn{3}{c}{Transition 1} & \multicolumn{3}{c}{Transition 2} \\
   \hline
   $i-j$   & Levels & WL ($\AA$) & $i-j$ & Levels & WL ($\AA$) \\
   \hline
   $20-21$ & $\mathrm{^6D_{9/2}-^6D_{7/2}}$ & $127\,345$ &
   $1-2$   & $\mathrm{^4F_{9/2}-^4F_{7/2}}$ & $84\,055$  \\
   $1-5$   & $\mathrm{^4F_{9/2}-^4P_{5/2}}$ & $5\,519.2$ &
   $4-7$   & $\mathrm{^4F_{3/2}-^4P_{1/2}}$ & $6\,121.0$ \\
   $1-26$  & $\mathrm{^4F_{9/2}-^4D_{5/2}}$ & $821.0$    &
   $1-25$  & $\mathrm{^4F_{9/2}-^4D_{7/2}}$ & $827.1$    \\
   $1-5$   & $\mathrm{^4F_{9/2}-^4P_{5/2}}$ & $5\,519.2$ &
   $20-25$ & $\mathrm{^6D_{9/2}-^4D_{7/2}}$ & $9\,524.8$ \\
   \hline
\end{tabular}

\flushleft{Key: WL, wavelength in vacuum, in $\AA$.}
\end{table}

The line intensity ratios are plotted in Figure \ref{fig:linrat} as a function 
of electron temperature and density.
The ratio between the lines $1-5$ and $20-25$ $(5\,519.2\,\AA)/(9\,524.8\,\AA)$
provides very powerful diagnostics for the electron temperature $T$.
It is density independent and varies significantly in the range of the peak 
abundance temperature.
The ratio $20-21/1-2$ $(127\,345\,\AA)/(84\,055\,\AA)$ similarly has a region 
where it is independent of density but the range is significantly greater then 
the temperature of maximum abundance for the $\mathrm{Ni}^{3+}$ ion. 
The ratio between lines $1-26$ and $1-25$ $(821.0\,\AA)/(827.1\,\AA)$
is a very useful density diagnostic particularly for low density plasmas, 
below $10^{9}\,\mathrm{cm^{-3}}$, in the range of the temperature of peak abundance.
Additional line ratios can be analysed using the present effective collision strengths
and with a more refined collision-radiative model.
We provide good-quality data to perform plasma modeling using \ion{Ni}{iv}
emission lines.

\begin{figure}
   \includegraphics[width=\columnwidth]{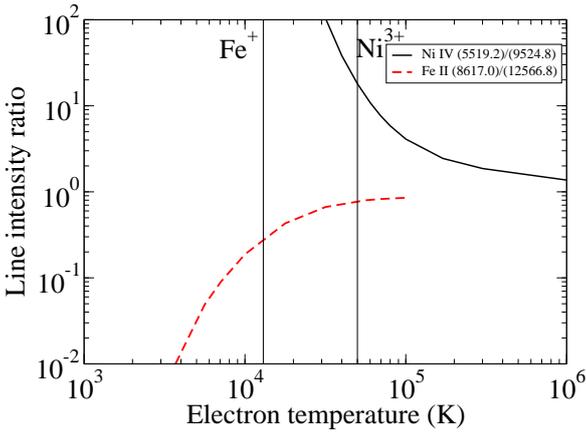} 
\caption{Line intensity ratio $\frac{I_{\lambda 1}}{I_{\lambda 2}}$ 
   versus electron temperature for a constant density 
   of $d=10^{4}\,\mathrm{cm}^{-3}$ for lines
   $(5\,519.2\,\AA)/(9\,524.8\,\AA)$ of \ion{Ni}{iv} (full line) and
   $(8\,617.0\,\AA)/(12\,566.8\,\AA)$ of \ion{Fe}{ii} (dashed line).
   Vertical lines indicate the peak abundance temperature of each ion.
   Colour online. }
\label{fig:linratcut}
\end{figure}

For \ion{Fe}{ii}, the equivalent wavelengths to the ratio between $1-5$ 
and $20-25$ are the lines of $8\,617.0$ and $12\,566.8\,\AA$.
The ratio of these lines can give a good diagnostics for the plasma temperature
if it is in the range of the peak abundance for $\mathrm{Fe}^{+}$
of $1.3 \times 10^{4} \kelvin$, see \citep{smyth2018b}.
Combining these two line ratios for \ion{Fe}{ii} and \ion{Ni}{iv} we are able
to determine with accuracy the electron temperature of the plasma in a wider range.
In figure \ref{fig:linratcut} we show the variation of these line ratios
for the electron density of the Orion nebula.

\section{Conclusions}
\label{sec:conclusions}

We present high-quality atomic data for electron-impact excitation 
of $\mathrm{Ni}^{3+}$ and photoionization from the ground and metastable levels 
of $\mathrm{Ni}^{2+}$.
These data are essential for the interpretation of $\ion{Ni}{iv}$ lines
collected from ground and satellite observations,
as well as opacity due to $\mathrm{Ni}^{2+}$ in interstellar clouds. 
A fully relativistic Dirac atomic $R$-matrix code ({\sc darc}) treatment is adopted
with a configuration interaction expansion of the 25-electron
target $\mathrm{Ni}^{3+}$ incorporating lowest 262 levels in the close coupling
expansion of the target. 
For each of the two processes we have performed two calculations, one using the 
calculated energies and atomic wave functions obtained within {\sc grasp}, 
and a second one replacing the calculated energies with the recommended
data tabulated in the NIST database. 
For both processes, the differences between the two calculations performed
was negligible with the background cross section as well as the height and 
positioning of the resonance structures almost identical in both.
Accuracy checks were performed throughout the analysis and we are confident 
that the present data represents the best available to date for use by the 
astrophysics and plasma physics communities. 

\section*{Acknowledgments}

Present work has been funded by the STFC through the 
QUB Astronomy Observation and Theory Consolidated Grant ST/P000312/1.
The computation has been performed in the supercomputer Hazelhen
property of the H\"ochstleistungsrechner f\"ur Wissenschaft und Wirtschaft (Germany),
and Archer property of the Engineering and Physical Science Research Council
under the allocation E464-RAMPA.

\bibliographystyle{mnras}
\bibliography{references}


\bsp	
\label{lastpage}
\end{document}